\documentclass[12pt]{article}
\usepackage{amsmath,amsthm}
\usepackage{psfrag}
\usepackage{graphicx}
\usepackage{amssymb,latexsym}

\newtheorem{theorem}{Theorem}[section]
\newtheorem{lemma}[theorem]{Lemma}
\newtheorem{statement}[theorem]{Statement}
\newtheorem{proposition}[theorem]{Proposition}
\newtheorem{corollary}[theorem]{Corollary} 
\newtheorem{definition}[theorem]{Definition} 
\theoremstyle{remark}
\newtheorem{remark}[theorem]{Remark}

{\hspace*{\fill}$\rule{.3\baselineskip}{.35\baselineskip}$\end{trivlist}}

\newcommand{\bi}{\begin{itemize}}
\newcommand{\ei}{\end{itemize}}
\newcommand{\bt}{\begin{theorem}}
\newcommand{\et}{\end{theorem}}
\newcommand{\bp}{\begin{proof}}
\newcommand{\ep}{\end{proof}}
\newcommand{\be}{\begin{equation}}
\newcommand{\ee}{\end{equation}}
\newcommand{\ben}{\begin{enumerate}}
\newcommand{\een}{\end{enumerate}}

\newcommand{\C}{\mathbb C}

\newcommand{\N}{\mathbb N}
\newcommand{\Z}{\mathbb Z}
\newcommand{\R}{\mathbb R}

\newcommand{\Bscr}{\mathcal B}
\newcommand{\Cscr}{\mathcal C}
\newcommand{\Dscr}{\mathcal D}

\newcommand{\Rscr}{\mathcal R}

\newcommand{\Sscr}{\mathcal S}

\newcommand{\hf}{\frac{1}{2}}

\newcommand{\e}{\varepsilon}

\newcommand{\m}{\mu}

\newcommand{\s}{\sigma}

\newcommand{\G}{\Gamma}

\renewcommand{\b}{\beta}

\newcommand{\z}{\zeta}
\newcommand{\x}{\xi}
\newcommand{\D}{\Delta}
\renewcommand{\S}{\Sigma}
\renewcommand{\a}{\alpha}
\renewcommand{\l}{\lambda}
\newcommand{\g}{\gamma}
\renewcommand{\d}{\delta}

\newcommand{\Res}{\text{Res }}
\renewcommand{\and}{\text{~~ and ~~}}
\renewcommand{\part}{\partial}
\newcommand{\ra}{\rightarrow}

\newcommand{\sign}{{\rm sign}}

\newcommand{\gt}{\hat \gamma}
\newcommand{\ie}{{i\over \e}}

\newcommand{\pt}{{\pi\over 2}}

\newcommand{\mt}{{\m\over 2}}

\newcommand{\sech}{{\rm sech~}}

\newcommand{\smf}{\sqrt{{\m^2\over 4}-1}}

\newcommand{\ipt}{{i\pi\over 2}}

\renewcommand{\D}{\Delta}

\newcommand{\msq}{{\mu^2\over 4}}

\begin{document}

\title{Semiclassical limit of the scattering transform for the focusing  Nonlinear Schr\" odinger Equation}

\author{Alexander Tovbis\footnote{
Department of Mathematics,
University of Central Florida,
Orlando, FL 32816, email: atovbis@pegasus.cc.ucf.edu~~~Supported by  NSF grant DMS 0508779}  \ and
Stephanos Venakides\footnote{
Department of Mathematics,
Duke University,
Durham, NC 27708, e-mail:
ven@math.duke.edu~~~
Supported by   NSF grant DMS
0707488}}

{\bf Abstract.} The semiclassical limit of the focusing Nonlinear (cubic) Schr\" odinger Equation (NLS) corresponds to the singularly perturbed Zakharov Shabat (ZS) system that defines the direct and
inverse scattering transforms (IST). In this paper, we derive explicit expressions for the leading 
order terms of these transforms, which are called semiclassical limits of 
the direct and inverse scattering transforms. Thus, we establish an explicit connection between the decaying initial data of the form $q(x,0)=A(x)e^{iS(x)}$ and the leading order term of its scattering data. This connection is expressed in terms of an integral transform that can be viewed as a
complexified version of an Abel type transform. Our technique is not based on the WKB analysys of the ZS system, but on the inversion of the modulation equations that solve the inverse scattering
problem in the leading order. The results are illustrated by a number of examples.

\maketitle
 
\section{Introduction}\label{intro}

Direct and inverse scattering  transforms 
play the key role in the solution of integrable
systems. Important results about the scattering transform with decaying initial data (potential)
can be found in \cite{Zhou}. Our main interest lies in  the semiclassical limit of 
the scattering transform for Zakharov - Shabat (ZS) system
\begin{equation}
\label{ZS}
i\e \frac{d}{dx} W=
\begin{pmatrix}
z & q\\
\bar q & -z
\end{pmatrix}W~,
\end{equation}
where $z$ is a spectral parameter, $\e$ is a small  positive parameter and $W$ is a $2$ by $2$ matrix-function. ZS system \eqref{ZS}
is the first (spatial) equation of the
Lax pair  for the  focusing Nonlinear Shr\"{o}dinger equation (NLS)
\be  \label{FNLS}
i\e\partial_t q + \hf\e^2\partial_x^2 q + |q|^2q=0 ~,\ \ \ \ \ \
\ee
where $x\in\R$ and $t\ge 0$ are space-time variables. 

The scattering data, corresponding to the initial data  $q(x,0,\e)$
consists of the reflection coefficient $r_{init}(z,\e)$,
as well as
of the points of discrete spectrum, if any, together with their norming constants. 
Since the time evolution of the scattering data is simple and very well known (\cite{ZS}),
the evolution of a given potential can be obtained through the inverse scattering of the
evolving scattering data. The inverse scattering problem for the NLS \eqref{FNLS} at
the point $x,t$ can be  
cast as a matrix Riemann-Hilbert Problem (RHP) in the spectral $z$-plane, which is stated as:
find a $2\times 2$ 
matrix-valued function $m(z)=m(z;x,t,\e)$, which depends on the asymptotic parameter $\e$ 
and the external parameters
$x,t$, such that: i) $m(z)$ is analytic in $\C\backslash \G$, 
where the contour $\G=\R$ with the natural orientation; ii)
\be\label{RHPmat}
m_+=m_-
\begin{pmatrix} 
1 + r\bar r  &  \bar r\cr
 r  &1 \cr
\end{pmatrix}
=m_- V~   
\ee
on the contour $\G$, where $r(z,\e)=r_{init}(z,\e) \exp [{2i\over \e}(2z^2 t +zx)]$
and $m_\pm(z)=\lim_{\d\ra 0}m(z\pm i\d)$ with $\d>0$ and $z\in\R$; iii)
$\lim_{z\ra\infty} m(z)=I,$
where $I$ denotes the identity matrix. 
In the presence of solitons 
 the contour $\G$  contains additional small circles around the eigenvalues 
with the corresponding jump-matrices (see, for example, \cite{TVZ1} or \cite{KMM}).

In a more general setting, corresponding to  AKNS systems, $\bar r(z)$ in the the jump matrix $V$ 
should be replaced by $\rho(z)$, which represents  another piece of the scattering data that 
is independent of $r(z)$;
 i.e., there is no functional dependence between $r(z)$ and $r^*(z)$ on $\R$; 
however, this case will not be considered
in the present paper.  
It is well-known (see, for example \cite{Zh})
that the 
RHP \eqref{RHPmat} has a unique solution $m(z)$ that has
asymptotics $m(z)=I+{m_1\over z}+O(z^{-2})$ as $z\ra\infty$, 
and that the solution to the NLS \eqref{FNLS} is given by  $q(x,t,\e)=-2(m_1)_{12}$,
where $(m_1)_{12}$ denotes the $(1,2)$ entry of matrix $m_1$. In the case when
 $r_{init}(z,\e)$ has analytic continuation into the upper halfplane, the RHP for $m(z)$ 
can be simplified by factorizing the jump matrix 
\be\label{fact1} 
V=
\begin{pmatrix} 
1 + rr^*  &  r^*\cr
 r  &1 \cr
\end{pmatrix}=\begin{pmatrix}
1 & \bar r\\
0 & 1
\end{pmatrix}
\begin{pmatrix}
1 & 0\\
 r & 1
\end{pmatrix}=V_-V_+~~,
\ee 
and ``splitting'' jump condition \eqref{RHPmat} into two jumps: one with triangular jump matrix
$V_+$ along some contour $\G_+$ in the upper halfplane $\bar \C^+$ ($\R$ is included in $\bar \C^+$)
 and the other
 with triangular jump matrix $V_-$ along some contour $\G_-$ in the  lower halfplane $\bar\C^-$.
Contours $\G_\pm$ are deformations of $\R$.
Due to the Schwarz symmetry of ZS problem, contours $\G_\pm$ can be choosen to be  symmetrical 
to each other with respect to the real axis, and we can restrict our attention to only
one jump condition, say, on the contour $\G_+\subset \bar\C^+$. 

A contour  $\G_+\in\bar\C^+$, which is a smooth deformation of $\R$, together with a function
$\tilde f(z)$, which is
analytic (or even H\"{o}lder
continuous) along  $\G_+$ with $\Im \tilde f(z)<-\d$ for all suffisiently large $z\in\G_+$, $\d>0$,
define a solution $\tilde q(x,t,\e)$ of the NLS \eqref{FNLS} in the following way:
if $\tilde m(z)$ is the normed at $z=\infty$ solution of the matrix RHP with the jump matrix
\be\label{rtilde}
V_+=\begin{pmatrix}
1 & 0\\
\tilde r & 1
\end{pmatrix},~~~~~~{\rm where}~~~~~\tilde r =e^{-\frac{2i}{\e}\tilde f(z)}
\ee 
on the contour $\G_+$, and the corresponding symmetrical  jump $V_-=V_+^*$ (see \eqref{fact1}) on the
symmetrical contour $\G_-$,
than 
\be\label{qt}
\tilde q(x,t,\e)=-2(\tilde m_1)_{12},~~~~~~{\rm where}~~~~~\tilde m(z)=I+{\tilde m_1\over z}+O(z^{-2})
\ee 
as $z\ra\infty$. This construction holds even if  $\tilde f(z)$  depends on $\e$.

In the semiclassical limit problem \eqref{FNLS}, we consider 
initial data (potential) of the form 
\be\label{ID}
q(x,0,\e)=A(x)e^{iS(x)/\e},
\ee
where
the amplitude  $A(x)$ is decaying at $\pm \infty$, and derivative of the phase 
$S'(x)$ has the limiting behavior
\be\label{s'inf}
\lim_{x\ra\pm\infty}S'(x)=\m_\pm~
\ee
with some finite $\m_-,\m_+, ~~m_-\le \m_+$.
In order to calculate the leading order of the solution $q(x,t,\e)$, $t\ge 0$, one needs
to find the leading order   of the solution $m$ to the RHP \eqref{RHPmat} as $\e\ra 0$.

The nonlinear steepest descent method (\cite{DZ1}, \cite{DZ2}), together with the 
$g$-function mechanism (\cite{DVZ}), is, perhaps, the most powerful tool of the asymptotic
analysis of the RHP \eqref{RHPmat}. The key part of this method, 
in the setting of our problem (genus zero region),  
is the following scalar RHP for the unknown  function $g(z)=g(z;x,t)$,  that: i)  is
 analytic (in $z$) in $\bar\C\setminus 
\g_m$ (including analyticity at $\infty$); ii) satisfies the jump condition
\be\label{rhpg}
g_+ + g_-=f_0-xz-2tz^2~~~~ {\rm on}~~\g_{m},
\ee
for  $x\in\R$ and $t\ge 0$, and; iii) has the endpoint behavior
\be\label{modeq}
g(z)=O(z-\a)^{3\over 2}~ + ~{\rm analytic~ function~ in~ a~ vicinity~ of~} \a. 
\ee
Here:  $\g_m$ is a  Schwarz-sym
 This observation
remains true ifmetrical contour
(called the main arc) with the endpoints $\bar\a, \a$, oriented from $\bar\a$ to $ \a$ and
intersecting $\R$ only at  $\m_+$; $g_\pm$ are the values of $g$ on the positive
(left) and negative (right) sides of $\g_m$, and;
function $f_0=f_0(z)$, representing the scattering
data, is   Schwarz-symmetrical and H\"{o}lder-continuous on $\g_m$.
Taking into the account Schwarz symmetry, it is clear
that  behavior of $g(z)$ at both endpoints $\a$ and  $\bar \a$ should be the same.
Assuming $f_0$ and $\g_m$
are known, solution $g$ to the RHP \eqref{rhpg} without the endpoint condition \eqref{modeq} 
can be obtained by Plemelj formula
\be\label{gform}
g(z)={{R(z)}\over{2\pi i}} \int_{\g_m}{{f(\z)}\over{(\z-z)R(\z)_+}}d\z~,
\ee
where
\be\label{fzxt}
 f(z)=f(z;x,t)=f_0(z)-xz-2tz^2
\ee 
and 
$R(z)=\sqrt{(z-\a)(z-\bar\a)}$. 
The branchcut of $R$ coinsides with $\g_m$ and the branch  of 
$R$ we use is defined by 
\be\label{brR}
\lim_{z\ra\infty} \frac{R(z)}{z}=-1~.
\ee
If $f_0(z)$ is analytic in some region $\Sscr$ that 
contains $\g_m\setminus\{\m_+\}$, the formula for $g(z)$   can be rewritten as
\be\label{gforman}
g(z)={{R(z)}\over{4\pi i}} \int_{\gt_m}{{f(\z)}\over{(\z-z)R(\z)_+}}d\z~,
\ee
where $\gt_m\subset\Sscr$ is a negatively oriented loop
around $\g_m$ (which is ``pinched'' to $\g_m$ in $\m_+$, where $f$ is not analytic)
that does not contain $z$.

It is well known (see, for example, \cite{Ga}) that additional smoothness at the endpoints
put some constrains on the location of these endpoints. To state these constrains, we introduce
 function $h=2g-f$. According to \eqref{gforman},  
\be\label{hform}
h(z)={{R(z)}\over{2\pi i}} \int_{\gt_m}{{f(\z)}\over{(\z-z)R(\z)_+}}d\z~,
\ee
where $z$ is inside the loop $\gt_m$. 
The endpoint condition
\eqref{modeq}  can now be written as
\be\label{modeqh}
h(z)=O(z-\a)^{3\over 2}~~{\rm as}~~z\ra \a,
\ee 
or, equivalently,
\be\label{modeqint}
\int_{\gt_m}{{f(\z)}\over{(\z-\a)R(\z)_+}}d\z=0~,
\ee
the latter equation known as a {\it modulation equation}. 
Substituting \eqref{fzxt} into \eqref{modeqint} yields
\be\label{xalpeq}
x+2(\Re\a+\a)t=\frac{1}{2\pi i}\int_{\gt_m}{{f_0(\z)}\over{(\z-\a)R(\z)_+}}d\z~.
\ee
It is clear that for a given  $f_0(z)$, \eqref{xalpeq}  defines $\a$ as a function $\a=\a(x,t)$.
The significance of $\a(x,t)$ is that  it represents the leading $\e$-order term
\be\label{q0}
q_0(x,t,\e)=A(x,t)e^{\ie S(x,t)}
\ee
of the solution $q(x,t,\e)$ to the Cauchy problem \eqref{FNLS}-\eqref{ID} through
\be\label{alpAS}
\a(x,t) = a(x,t)+ib(x,t)=-\hf S_x(x,t)+iA(x,t)~.
\ee

By  Schwarz symmetry, modulation equation \eqref{modeqint} holds if $\a$ is replaced by $\bar\a$.
Adding and subtracting these two equations and using integration by parts, we obtain
system of {\it moment conditions} 
\be\label{momloop}
\int_{\gt_m}{{f'(\z)}\over{R(\z)_+}}d\z=0~,~~~~~~\int_{\gt_m}{{(\z-a)f'(\z)}\over{R(\z)_+}}d\z=0~,
\ee
where $\a=a+ib$, which is  equivalent to \eqref{modeqint}. 
If $f_0$ is defined only on the contour $\g_m$ (nonanalytic case), the loop integrals
in \eqref{momloop} should be replaced by the integrals over the contour $\g_m$.
Substituting \eqref{fzxt} into \eqref{momloop}, we obtain
\be\label{momloopxt}
\frac{1}{2\pi i}\int_{\gt_m}{{f'_0(\z)}\over{R(\z)_+}}d\z=x+4ta~,
~~~~~~\frac{1}{2\pi i}\int_{\gt_m}{{(\z-a)f'_0(\z)}\over{R(\z)_+}}d\z=-2tb^2~.
\ee
For a fixed $t\ge 0$, the second moment equation \eqref{momloopxt} defines a curve $\S$ in the spectral
plane, whereas the first moment equation \eqref{momloopxt} determines a parametrization of $\S$ by $x\in\R$.

Solving system \eqref{momloopxt} for a given $f_0(z)$, i.e., finding $\a(x,t)$ that satisfies \eqref{momloopxt} for all $x\in\R$ and all $t\in[0,t_0]$ with some $t_0>0$, is the central part of the inverse scattering procedure  for the leading order solution of the Cauchy problem \eqref{FNLS}-\eqref{ID}. Considerable progress has been achieved in solving this problem, 
see \cite{TVZ1}, \cite{KMM}, \cite{TVZ3}. However, calculating $f_0$, which represents the leading
order term of the spectral data corresponding to \eqref{ID}, continue to pose a considerable challenge.
In particular, the spectral data considered in \cite{TVZ1} and \cite{KMM} was calculated explicitly,
because ZS system \eqref{ZS} for the corresponding initial data  was reduced to 
the hypergeometric equation. In the case of general analytic initial data \eqref{ID}, 
the WKB analysis of singularly perturbed ZS systems \eqref{ZS}
in the complex $x$-plane seems to be the most  natural approach
for the direct scattering. There are, however, considerable difficulties associated with this 
approach  even for relatively simple initial data, such as, for example, the need to keep track
of a large number of turning points, singularities, Stokes lines that connect them, etc.,  
(see, for examle, \cite{Mil}). This is why, in our opinion, the results about the direct scattering 
are quite limited: one can mention  numerical simulations of
the discrete spectrum of \eqref{ZS} in \cite{Bronski1}, followed by formal WKB calculation in \cite{Mil} of the $Y$-shaped spectral curve from \cite{Bronski1}, and  rigorous WKB construction of discrete spectrum
for certain special potentials \eqref{ID} in \cite{ST}. 

The main goal of the present paper is to derive  an explicit formula for $f_0(z)$ that will 
be valid for a rather broad class 
of initial data \eqref{ID} directly from the moment conditions \eqref{momloop} 
(Section \ref{AbelHilbert}). We proceed with studying properties of the transforms that connect
$f_0(z)$ with the scattering data (Sections \ref{inversion} - \ref{symmetry}) and,
at the end, consider a number of examples that include already studied potentials \eqref{ID}, 
as well as some new cases (Section \ref{examples}).  

For the rest of the paper, unless specified otherwise, we assume $t=0$.  Then the first equation
of \eqref{momloopxt} can be considered as a transformation 
\be\label{ftox}
x(\a)=\frac{1}{\pi i}\int_{\g_m}{{f'_0(\z)}\over{\sqrt{(\z-\a)(\z-\bar\a)}_+}}d\z
\ee
of a given $f'_0(\z)$ into  $x(\a)$, which has the meaning of the inverse function to 
$\a(x)=\a(x,0),~~x\in\R$. Provided that $\a(x)$, determined by initial data \eqref{ID} through  \eqref{alpAS}, is invertible for all  $x\in\R$,  
$x(\a)$ is a real valued function defined on the curve $\S$ that is the graph 
of $\a(x), ~~x\in\R$.  The main result of this paper is the formula
\be\label{xtof}
f_0(z)=\int^{\m_+}_{z}\left[z-\m_+ + \sqrt{(z-u)(z-\bar u)} \right]x'(u)du +(z-\m_+) x(z)+ f_0(\m_+),
\ee
where $z\in\S$ and the integral is taken along $\S$,
which is {\it the inversion} of transformation \eqref{ftox}.  
Here $f_0(\m_+)$ is a free real parameter and the branch of the radical satisfies  normalization \eqref{brR}. 
Transformations \eqref{ftox} and \eqref{xtof} resemble the   pair of 
{\it Abel transformants} for axially symmetric functions, which is convenient to write in the form 
(\cite {Bracewell})
\be\label{Abel} 
M(\x)=-2\int_\x^\infty N'(\eta)\sqrt{\eta^2-\x^2}d\eta~~~~~{\rm and}~~~~~~
N(\eta)=-\frac{1}{\pi}\int_\eta^\infty M'(\x)\frac{d\x}{\sqrt{\x^2-\eta^2}}~,
\ee
where $\x,\eta \in\R$. Indeed, if $\a\in i\R$ and $\S\subset i\R$, the radicals in  \eqref{ftox} and \eqref{xtof}
become $\sqrt{\z^2-\a^2}$ and $\sqrt{z^2-u^2}$ respectively, and transforms 
 \eqref{ftox} and \eqref{xtof} become a pair of Abel transforms (with 
some extra terms
in \eqref{xtof} required for  convergence). However, if $\a\in \R$ and $\S\subset \R$,
 \eqref{ftox} becomes  finite Hilbert transform (note that, due to Schwarz symmetry, 
$f'_0(\z)$ has a jump $2i\Im f'_0(\z)$ along the real axis) on $[a,\m_+]$.
Transformations \eqref{ftox} and \eqref{xtof} will be referred to as  {\it Abel-Hilbert (AH) or
complexified  Abel transformations}  defined on a contour $\S\subset \C$. 

A given initial data \eqref{ID} determines $\a(x)=\a(x,0)$ by \eqref{alpAS}, where $A(x,0)=A(x)$
and $S_x(x,0)=S'(x)$. We consider the following three objects, defined by initial data \eqref{ID}:
 1) solution $q(x,t,\e)$ of the 
Cauchy problem \eqref{FNLS}, \eqref{ID}; 2) solution $\tilde q(x,t,\e)$  to \eqref{FNLS} defined
through \eqref{rtilde}-\eqref{qt}, where $\tilde f=f_0(z)-xz-2tz^2$ with $f_0(z)$
being the AH transformation of $x(\a)$ given
by \eqref{xtof}, and $\G_+=\S$ (here we assume that $\a(x)$, $x\in\R$, is invertible with the
inverse $x(\a)$; 3) function $q_0(x,t,\e)$ that is defined by \eqref{q0}, \eqref{alpAS}, where 
$\a(x,t)$ is a smooth in $x\in\R$ and $t\ge 0$   solution of the moment conditions  \eqref{momloopxt}
(analyticity of the initial data \eqref{ID} on $x\in\R$ is required to define $q_0(x,t,\e)$ with
$t>0$).
Since  $q(x,0,\e)=q_0(x,0,\e)$  (transformation \eqref{ftox} is inverse to \eqref{ftox}), we call
$q_0(x,t,\e)$  a {\it leading order semiclassical solution} to Cauchy problem \eqref{FNLS}, 
\eqref{ID}, or simply a {\it  semiclassical solution}. The main achievement of this paper is 
formal construction of the semiclassial solution $q_0(x,t,\e)$ with some positive $t$ for a given
analytic initial data \eqref{ID}.   
Does the semiclassical solution $q_0(x,t,\e)$ indeed represent a leading order 
behavior of $q(x,t,\e)$ for $t>0$? Although we do not have the general answer to this question,
the following facts and observations indicate that the answer should be affirmative at least for
some substantial class of the analytic initial data \eqref{ID}. 

Requirements guaranteeing that $q_0(x,t,\e)$ is $O(\e)$ close to $\tilde q(x,t,\e)$ 
on a compact subset $D$ of the $x,t$ plane that
do not contain breaking points (see below), can be found in
\cite{TVZ1}, \cite{TVZ3}. The first requirement is that 
\be\label{ineq2}
w(z)=\sign(\m_+-z) \Im f_0(z)<0~~~{\rm for}~~~ z<\m_-~~~~~{\rm and~~ for}~~~~~ z>\m_+;
\ee
more precisely, $w(z)$ is separated from zero if $ z<\m_-$ and if $z>\m_+$ except 
for neighborhoods of $\m_\pm$, where it has behavior $O(z-\m_\pm)$ respectively.
Let $\g_{c}$ be a bounded smooth oriented contour in the upper halfplane, called complementary arc, 
that connects $\a$ and $\m_-$ and does not intersect $\g_m$ (except at the common endpoint $\a$).
The second requirement is that for any $(x,t)\in D$ there exist main and complementary
arcs  $\g_m$ and $\g_c$, connecting $\a(x,t)$ with $\m_\pm$ respectively, such that  
the signs of $\Im h(z)$, where $h$ is  defined by \eqref{hform},
satisfy inequlities (sign distributions)
\be\label{ineq1}
\Im h(z)<0~~~{\rm on ~both~ sides~ of}~\g_m~~~~{\rm and} ~~~~\Im h(z)>0~~~{\rm on~ at ~ least~ one~ side ~of} \g_c. 
\ee

As shown in Section \ref{break},  violation of smoothness of $q_0(x,t,\e)$ in the process of
time evolution leads to the break of the anzatz \eqref{q0} for the semiclassical solution.
Since \eqref{ineq2} is independent of $x,t$, the break at  some point $ (x_b,t_b)$ means that 
at least one of the inequalities \eqref{ineq1} is violated at some point(s) $z_b\in\g_m\cup\g_c$, 
$z_b\neq \a(x_b,t_b)$, or that condition \eqref{modeqh} becomes
\be\label{breakatalph}
h(z)=o(z-\a)^{3\over 2}~~{\rm as}~~z\ra \a,
\ee 
if $z_b=\a(x_b,t_b)$. If $z_b$ is not a branchpoint of $f_0(z)$ (regular break), 
then the situation can be corrected
by introducing additional main and complementary arcs in the RHP \eqref{rhpg}-\eqref{modeq},
see \cite{TVZ1} for details. This corresponds to the change of genus from zero to a positive even number
(the genus is even because of Schwarz symmetry) of some hyperelliptic
Riemann surface $\Rscr(x,t)$  that shadows the evolution of $\tilde q(x,t,\e)$,
see Remark \ref{higherg} below. Correspondingly,
the semiclassical solution $q_0(x,t,\e)$ can be expressed in terms of Riemann theta functions
defined by $\Rscr(x,t)$, and  $O(\e)$ closeness between $q_0(x,t,\e)$ and $\tilde q(x,t,\e)$ 
 extends to the region beyond the break. $O(\e)$ accurate approximation of   $\tilde q(x,t,\e)$ 
by $q_0(x,t,\e)$ can be extended through further breaks (see \cite{TV3}) provided that the breaks
are regular. The case when $z_b$ is a branchpoint of $f_0(z)$  (singular break) requires additional
study. Approximation of   $\tilde q(x,t,\e)$ by $q_0(x,t,\e)$ allows us to call 
$q_0(x,0,\e)$ the {\it semiclassical limit of the initial data} for solution $\tilde q(x,t,\e)$
(which is determined by $f_0(z)$).

Does $q_0$ approximate solution $q$ of the Cauchy problem \eqref{FNLS}, \eqref{ID}? The answer 
to this question depends on how accurately solution $\tilde q$ approximates   solutions $q$.
A closely related question is how accurately  the scattering data 
$e^{-\frac{2i}{\e}\tilde f_0(z)}$ (see \eqref{rtilde}) of $\tilde q$ approximates the scattering data
$r_{init}(z,\e)$ of $q$. The authors are not aware of any general results of this nature,
however, for a special family of initial data where explicit form of $r_{init}(z)$ is available (see \cite{TVZ1}),
\be\label{f0limfinit}
f_0(z)=\lim_{\e ra 0} \hf i\e\ln r_{init}(z,\e)
\ee
at every $z$ in the domain of analyticity of $\ln r_{init}(z,\e)$ (with properly located 
 branchcuts of $f_0(z)$). Similar result was obtained
in \cite{KMM} for  pure soliton solutions of \eqref{FNLS} ($r_{init}\equiv 0$), where 
$e^{-\frac{2i}{\e}\tilde f_0(z)}$ provided a good approximation to the corresponding discrete
scattering data. The authors expect that \eqref{f0limfinit} holds for a wide class of general
analytic initial data. That is why $f_0(z)$, obtained by AH transformation \eqref{xtof},
is called the {\it   semiclassical limit of the scattering data} that corresponds to \eqref{ID}.
The authors also expect that, subject to certain requirements, the semiclassical solution $q_0$
is  the leading order approximation of $q$ as $\e\ra 0$.
Establishing this fact seems to be 
the last remaining essential step towards the complete solution of the semiclassical asymptotic problem
for the focusing NLS.

\begin{remark}\label{higherg}
The RHP \eqref{rhpg} can be modified 
to include $N$, $N\in\N$, contours (main arcs) where $g$ undergoes a jump. These contours
define a hyperelliptic Riemann surface 
$\Rscr=\Rscr(x,t)$ of the genus $N-1$, associated with the semiclassical limit of the initial value proble (Cauchy problem) \eqref{FNLS}-\eqref{ID}. Higher genus of $\Rscr$ indicates that  
the corresponding potential can be represented as a modulated $N$-phase wave expressed through 
the Riemann theta-functions, see, for example, \cite{TVZ1}. The authors belive that the 
AH transformation \eqref{xtof} can be generalized to reperesent the 
semiclassical limit of the direct scattering transform  for the higher genus
cases, however, this paper is restricted to study  potentials of the form \eqref{ID} only, i.e.,
to genus zero potentials.
\end{remark}

\section{Inversion formula for the AH transform} \label{AbelHilbert}

In this section we prove that under the appropriate assumptions on $\a(x)$ the transformation
\eqref{ftox} inverts the transformation \eqref{xtof}. Let us assume that:

\begin{enumerate}\label{umm}
 \item $\a(x)=-\hf S'(x) + iA(x)$ is a complex valued $C^1$ function on $\R$, where $A(x)$ is positive
and  $S'(x)$ satisfies \eqref{s'inf};
 \item $\a(x)$ is locally
and globally invertible, i.e., $\a'(x)\neq 0$ on $\R$ and the graph $S$ of $\a(x)$ does not
have points of self-intersection;
 \item the inverse function $x(\a)$, $\a\in\S$, satisfies
\be\label{xliminf}
\lim_{u\ra \m_+} (u-\m_+)x(u) =0,~~u\in\S,~~~~~{\rm and}~~~~~~(u-\m_+)x'(u)\in L^1(\S_{u_0})
\ee
for any $u_0\in\S$, where $\S_{u_0}$ denotes the arc of $\S$ connecting $u_0$ and $\m_+$.
\end{enumerate}

\begin{theorem}\label{invtheo}
If $\a(x)$ satisfies conditions \ref{umm} then $f_0(z)$ in \eqref{xtof} and its derivative
are well defined. Moreover,
transformation \eqref{ftox} is inverse to 
transformation \eqref{xtof}, i.e., the substitution of \eqref{xtof} into \eqref{ftox} turns
the latter one into the identity.
\end{theorem}

\bp
According to our choice \eqref{brR} of the branch of the radical in \eqref{xtof}, we 
have 
\be\label{radlimm+}
\sqrt{(z-u)(z-\bar u)} \sim -(z-\m_+)
\ee
as $u\ra\m_+$ provided $z$ is separeted from $\m_+$. Then
\be\label{convf}
z-\m_+ + \sqrt{(z-u)(z-\bar u)}=-\frac{2(z-\m_+)(\m_+-\Re u)+|\m_+-u|^2}
{\left[ z-\m_+ - \sqrt{(z-u)(z-\bar u)}\right] \sqrt{(z-u)(z-\bar u)}}~,
\ee
where, according to \eqref{radlimm+}, the denominator approaches $-2(z-\m_+)^2$ as $u\ra\m_+$.
Now, convergence of the integral in \eqref{xtof} follows from \eqref{convf} and the second condition
of \eqref{xliminf}.
Differentiation of \eqref{xtof} yields
\be\label{xtof'}
f'_0(z)=\int^{\m_+}_{z}\left[1 + \frac{z-\Re u}{\sqrt{(z-u)(z-\bar u)}} \right]x'(u)du + x(z),
\ee
where the integrand can be expressed as
\be\label{convf'}
1 +\frac{z-\Re u}{\sqrt{(z-u)(z-\bar u)}}=-\frac{(\Im u)^2}
{\left[ z-\m_+ - \sqrt{(z-u)(z-\bar u)}\right] \sqrt{(z-u)(z-\bar u)}}~.
\ee
Since $|\Im u| < |u-\m_+|$, convergence of the integral in \eqref{xtof'} follows from \eqref{convf'} 
and the second condition of \eqref{xliminf}.

Substituting of \eqref{xtof'} into \eqref{ftox}, which can be converted into
\be\label{ftoxim}
x(\a)=\frac{2}{\pi }\Im \int_{\m_+}^\a{{f'_0(\z)}\over{\sqrt{(\z-\a)(\z-\bar\a)}_+}}d\z,
\ee
yields
\begin{align}\label{inver1}
x(\a)=\frac{2}{\pi }\Im \int_{\m_+}^\a\left[ \int_\z^{\m_+}\frac{1+ \frac{z-\Re u}{\sqrt{(z-u)(z-\bar u)}}}
{\sqrt{(\z-\a)(\z-\bar\a)}}x'(u)du + \frac{x(\z)}{\sqrt{(\z-\a)(\z-\bar\a)}}\right] d\z=&\cr
=-\frac{2}{\pi }\Im \int^{\m_+}_\a\left[x'(u)\int_\a^u\frac{1+ \frac{z-\Re u}{\sqrt{(z-u)(z-\bar u)}}}
{\sqrt{(\z-\a)(\z-\bar\a)}}d\z+\frac{x(u)}{\sqrt{(u-\a)(u-\bar\a)}}\right] du&\cr~.
\end{align}

\begin{figure}
\centerline{
\includegraphics[height=9cm]{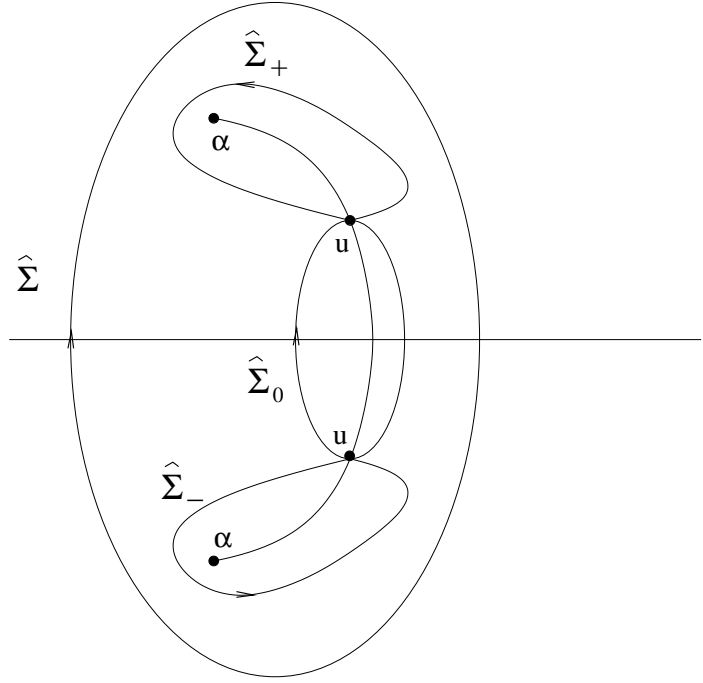}}
\caption{Contours  $\hat\S$, $\hat\S_0$ and $\hat\S_\pm$. }
\label{AHconts}
\end{figure}

Denote by
\be\label{pz}
p(\z)=\frac{1+ \frac{z-\Re u}{\sqrt{(z-u)(z-\bar u)}}}
{\sqrt{(\z-\a)(\z-\bar\a)}}
\ee 
and by $\hat\S$ a negatively oriented loop that contains contour $\S_\a\cup\overline{\S_\a}$. Then 
$\oint_{\hat\S}p(\z)d\z=0$ by Residue Theorem. If $\hat\S_0$ denotes a negatively oriented loop
around contour $\S_u\cup\overline{\S_u}$, and $\hat\S_\pm$ denote positively oriented loops around 
contour $\S_\a\setminus \S_u$ and its complex conjugate respectively, see Figure \ref{AHconts}, then
\be\label{pzident}
\oint_{\hat\S_0}p(\z)d\z=\oint_{\hat\S_+}p(\z)d\z+\oint_{\hat\S_-}p(\z)d\z~.
\ee 
However, $\oint_{\hat\S_0}p(\z)d\z=\oint_{\hat\S_0}\frac{d\z}{\sqrt{(\z-\a)(\z-\bar\a)}}$,
since the remaining term of $p(\z)$ attains  the same values on both sides of the branchcut 
$\S_u\cup\overline{\S_u}$. Taking into the account Schwarz symmetry of the intgrands, we obtain
\be\label{pzend}
\Im \int_\a^u p(\z) d\z = \Im \int_{\m_+}^u \frac{d\z}{\sqrt{(\z-\a)(\z-\bar\a)}}~.
\ee

Returning to \eqref{inver1} and taking into the account \eqref{pzend} and the 
that $d x(u)=x'(u)du\in\R$ when $u\in\S$, we obtain
\begin{align}\label{Iminout}
\Im \int^{\m_+}_\a\left[x'(u)\int_\a^u p(\z)d\z\right]du= & \int^{\m_+}_\a x'(u)\Im \left[\int_\a^u p(\z)d\z\right]du =\cr 
\int^{\m_+}_\a x'(u)\Im \left[ \int_{\m_+}^u \frac{d\z}{\sqrt{(\z-\a)(\z-\bar\a)}}\right]du = &
\Im\int^{\m_+}_\a x'(u) \int_{\m_+}^u \frac{d\z}{\sqrt{(\z-\a)(\z-\bar\a)}}du~.
\end{align}
Thus,
\begin{align}\label{inver2}
x(\a)=-\frac{2}{\pi }\Im \int_{\m_+}^\a\left[x'(u)\int_{\m_+}^u \frac{d\z}{\sqrt{(\z-\a)(\z-\bar\a)}}+
\frac{x(u)}{\sqrt{(u-\a)(u-\bar\a)}}\right] du=& \cr
-\frac{2}{\pi }\left. x(u)\Im \int_{\m_+}^u \frac{d\z}
{\sqrt{(\z-\a)(\z-\bar\a)}}\right|^{u=\m_+}_{u=\a}=
x(\a)\left(\frac{1}{2\pi i} \int_{\hat S} \frac{d\z}{\sqrt{(\z-\a)(\z-\bar\a)}}\right) &=x(\a)~,
\end{align}
where we used the fact that, according to the limit in \eqref{xliminf}, the contribution from $u=\m_+$ is zero.
\ep

\section{Derivation of transformation \eqref{xtof} for $f_0(z)$} \label{inversion}

So far, the observation that transformation \eqref{ftox} resembles Abel transformation helped us
to guess transformation \eqref{xtof}, to which \eqref{ftox} is inverse. In this section, we will 
show how transformation \eqref{xtof} can be derived from the analysis of the RHP \eqref{rhpg}.
In particular, we show that solution $g(z)$ for the  RHP \eqref{rhpg}, represented by 
integral \eqref{gforman}  in  the complex  $z$-plane (Plemelj formula), can  also be represented by 
a dual  integral in the  complex $x$-pane
(see \eqref{gxzfin} below). The ``input'' data for the integral representation 
in the $z$-plane is $f_0(z)$ and $x$,
whereas the ``input'' data for the dual integral representation 
in the $x$-plane is $\a(x)$ and $z$. 
Function $h(z)$ has similar integral representations. Then teh semiclassical limit of the spectral data
$f_0(z)$ is given by $f_0=2g-h +xz$.
 
We start our derivation with the following observation.

\begin{proposition}\label{dRdalp=0} 
Conditions \eqref{rhpg}-\eqref{modeq} imply that $\frac{\part g(z)}{\part \a}\equiv 0$.
\end{proposition}

\bp
Applying
\be\label{dRdalp}
\frac{\part R(z)}{\part \a}=-\frac{R(z)}{2(z-\a)}~
\ee
and  $\frac{1}{(\z-z)(\z-\a)}=\frac{1}{z-\a}\left[\frac{1}{\z-z}-\frac{1}{\z-\a}\right] $ to \eqref{gform}, we obtain 
\be\label{dRdalpint}
\frac{\part g(z)}{\part \a}={{R(z)}\over{8\pi i(z-\a)}} \int_{\gt_m}{{f(\z)}\over{(\z-\a)R(\z)_+}}d\z.
\ee
But, according to \eqref{modeqint}, the integral in \eqref{dRdalpint} is zero.
The proof is completed.
\ep 

Under our convention $t=0$,  functions $R, g,h,f$ depends on $x$ and $z$.
Since $z$ will be considered as a parameter for the rest of the paper (unless specified
otherwise), it is convenient for us henceforth to put the variable $x$ in these functions 
in the first position. For example, the radical $R$ introduced in  \eqref{gform}
can be now rewritten as
\eqref{R} 
\be\label{R}
R(x,z)=\sqrt{(z-a(x))^2+b^2(x)}~
\ee
with the same choice of the branchcut in the $z$ plane as before.
The choice of the branchcut for $R(x,z)$ in the complex $x$-plane is discussed below. 

Since $f(x,z)=f_0(z)-xz$, Proposition \ref{dRdalp=0} implies that total derivatives
\be\label{dh/dxt}
\frac{d}{dx}g(x,z)\equiv \frac{\part }{\part x}g(x,z),~~~~~~~{\rm and}~~~~~~~~~
\frac{d}{dx}h(x,z)\equiv \frac{\part }{\part x}h(x,z)~
\ee
coincide with the corresponding partial derivatives. 
Differentiating both sides of the RHP \eqref{rhpg} in $x$ and using \eqref{modeq}, we obtain
\be\label{gh_x}
\frac{d h(x,z)}{d x}= -R(x,z),~~~~\frac{d g(x,z)}{d x}= -\hf[z+R(x,z)]
\ee
for all $x\in \R$ and $z\in \overline{\C^+}$, where $z$ is considered as a parameter 
and $\C^\pm$ denotes the upper and the lower halfplane respectively
(we also use notation $\cal B^\pm=\cal B\cap \C^\pm$ for any set $\cal B$). 
The  fact that derivatives in \eqref{gh_x} are independent of $f$ opens the way
to reconstruct $h(x,z),~g(x,z)$ and, thus, $f_0(z)$.

To construct $f_0(z)$, we require that  
\be\label{alph}
 \a(x)=a(x)+ib(x)= -\hf S'(x)+iA(x)
\ee 
satisfies the following conditions {\bf (A)} :
\begin{enumerate}\label{reqab}

\item $a(x)$ and $b(x)$ are real analytic on $\R$ and $b(x)> 0$;
\item all but finitely many points of the parametric curve
\begin{equation} \label{Edef}
\partial {\cal E}=\{z \in \C:\; z=-S'(x)/2\pm i A(x),\; x \in \R\}
\end{equation}
are regular points, i.e., the tangent vector $\a'(x)\ne 0$, $x\in\R$; moreover,
$\partial {\cal E}$ does not have points of self-intersection (note that we used notation $\S$ 
for $\partial {\cal E}^+$ in Sections \ref{intro}, \ref{AbelHilbert});
\item if $\Dscr $ denotes the common domain of analyticity of $a(x),b(x)$, then
$\a(\Dscr)\supset{\cal L}$; here $\a(x)$ is defined by \eqref{alph}
and ${\cal L}$ is an open domain in the upper $z$-halfplane that contains  
the union of the strip $\{z:~0\le \Im z<L\}$, where $L>0$, with 
$\overline{{\cal E}}^+$, where $\cal E$ denotes the open region of $\C$ bounded by $\part{\cal E}$; 
%
\item 
there exist $\m_\pm\in\R$ and $p>1$ such that 
\be\label{ablim}
\lim_{\Re x\ra\pm\infty}x^p\left[ a(x)-\m_\pm\right] = \lim_{\Re x\ra\pm\infty}x^pb(x)=0,
\ee 
respectively for all $x$ such that $\a(x)\in{\cal L}$;  

\end{enumerate}

The curve $\partial {\cal E}$
 connects $\m_+$ and $\m_-$ in both $\C^+$ and $\C^-$.  
Condition {\bf A2} implies that there exists the inverse function $x(z)$, where 
$x(\a(x))\equiv x$~~ $\forall x\in\R$. Note that $x(z)$ is analytic on 
$\part{\cal E}^+$ at all the regular points. 
Any point $x^*\in\Dscr$, such that
$\a'(x^*)=0$, is called a point of ramification. The corresponding $z$ is called
a logarithmic (log) point. We can analytically continue $x(z)$ from 
$\part{\cal E}
\begin{figure}
\centerline{
\includegraphics[height=6cm]{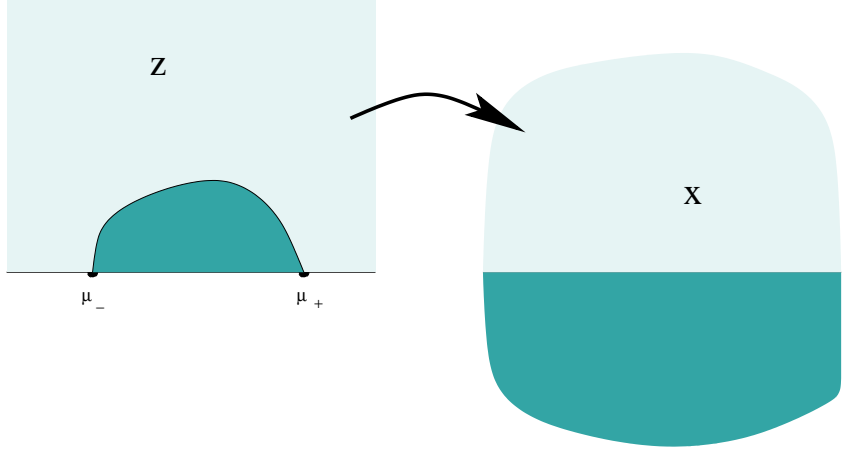}}
\caption{Map $x(z)$ maps $\part{\cal E^+}$ into the real $x$-axis, $\overline{\cal E^+}$  (darker area)
into $\Bscr^-$ and the rest of ${\cal L}$ (lighter area) into $\Bscr^+$. Possible cuts
are not shown here.)}
\label{xmap}
\end{figure}

^+$ to ${\cal L}$ with the exception of the log points.   To define $x(z)$ uniquely
in, ${\cal L}$
we make branchcuts connecting  every log point  $z^*\in{\cal L}, ~\Im z>0$ with $\R\cup{\infty}$ in such a way that they do not intersect ${\cal E}^+$ and keep ${\cal L}$ connected. 
Let $\Cscr$ denote ${\cal L}$
with the cuts.
Then $\Bscr=x(\Cscr)$ is the image of $\Cscr$ by the conformal map $x(z)$, see
Fig. \ref{xmap}.    

\begin{figure}
\centerline{
\includegraphics[height=3cm]{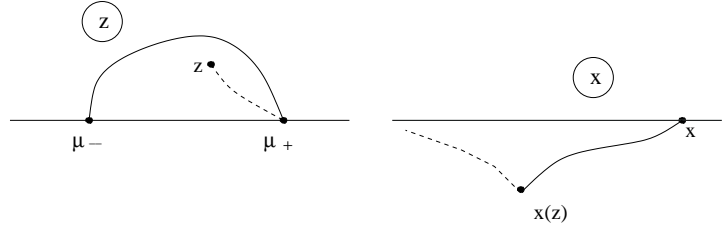}}
\caption{The branchcut of $R(x,z)$ is shown by the dashed lines:  a path connecting $\a(x)$ and $\m_+$,~~
$x$ is fixed (left); a path   connecting $x$ and $-\infty$,~~ $z$ is fixed (right).  }
\label{xcut}
\end{figure}

According to \eqref{gh_x}, we have 
\be \label{hxz}
h(x,z)=-\int_{x(z)}^x R(y,z)dy
\ee
for all $z\in\Cscr$ and all $x\in \Bscr$, where the contour of integration 
lies in $\Bscr$ and does not cross the branchcut $\G(z)$ of $R(x,z)$ in the complex $x$-plane ($z$ is fixed),
which is a path connecting $x(z)$ with $-\infty$
 (see Fig. \ref{xcut}). 

For any fixed $z\in\Cscr$,  function $h(x,z)$ is analytic in $x$ for all finite $x\in{\cal B}$
except the branchpoints $x=x(z)$ and $x=\overline{x(\bar z)}$ of the radical $R$ (the latter may or may not be in ${\cal B}$. On the other hand, 
\be\label{h_z0}
\frac{\part h}{\part z} (x,z)=- \int_{x(z)}^x \frac{z-a(y)} {\sqrt{(z-\a(y))(z-\tilde\a(y))}}dy+
\frac{\sqrt{(z-\a(x(z)))(z-\tilde\a(x(z)))}}{\a'(x(z))} ~,
\ee
where
\be\label{alphat}
\tilde\a(x)=\overline{\a(\bar x)}=a(x)-ib(x)~.
\ee
Since $\a(x(z))\equiv z$ in $\Bscr$, it is clear that the latter term is zero
for all $z$ except when
$\a'(x(z))=0$, i.e., except when $z=\m_\pm$ or $z$ is a log point. So,
\be\label{h_z}
\frac{\part h}{\part z} (x,z)=- \int_{x(z)}^x \frac{z-a(y)} {\sqrt{(z-\a(y))(z-\tilde\a(y))}}dy ~
\ee
if $\a'(x(z))\neq 0$. That means that for any
finite
$x\in\Bscr$ function $h(x,z)$ is analytic in $z\in\Cscr$ except for the log points,
the branchpoints
$z=\a(x)$, $z=\tilde \a(x)$, 
and, possibly,  points $z=\m_\pm$.

Does  $h(x,z)$ from \eqref{hxz} coincide with $h$ given by \eqref{hform}? If the corresponding 
$\a(x)$ and $f_0(z)$ satisfy
 modulation equation \eqref{modeqint}, then they may differ
by an independent of $x$ constant. However, this constant is identically zero since 
$h(x(z),z)\equiv 0$ for all $z\in{\cal C}$
in both cases of  \eqref{hxz} and \eqref{hform}.
 Thus, for any fixed $x\in{\cal B}$,
\be\label{hsamodeq}
h(x,z)=O(z-\a(x))^{\frac{3}{2}}~~~~{\rm as}~~~~z\ra\a(x), 
\ee
provided that $\a'(x)\ne 0$.
Equation \eqref{hsamodeq} also follows directly from \eqref{h_z}. 
Finally, for all $x\in\R$, we extend $h(x,z)$ into the lower $z$-halfplane 
$\C^-$ by Schwarz reflection. Note that
$h(x,z)$ has a jump $2i\Im h(x,z)$ for $z\in\R$.  In the case $z\in\C^-$, the values of  
$h(x,z)$ can be analytically continued from $x\in\R$ to the complex $x$-plane.

To calculate  $g(x,z)$, we first note that, according to \eqref{R}, \eqref{brR} and condition 
{\bf A3},
\be \label{limR}
R(x,z)=-(z-\m_\pm)+o\left( x^{-p}\right)~~~~{\rm as}~~~  \Re x\ra \pm\infty,~~x\in{\cal D}  
\ee
for any fixed $z\in\C$. Then
\begin{align}\label{limzmR}
z-\m_\pm +R(x,z)=&\frac{2(z-\m_\pm)[a(x)-\m_\pm]-[a(x)-\m_\pm]^2-b^2(x)}{z-\m_\pm-R(x,z)}=\cr
&a(x)-\m_\pm-\frac{R^2(x,\m_\pm)}{2(z-\m_\pm)}+o(x^{-2p})
\end{align}
as $\Re x\ra \pm\infty,~x\in{\cal D}$, for any fixed $z\neq \m_\pm$ respectively.
According to  \eqref{gh_x}, we define $g(x,z)$ as
\be\label{gxz}
g(x,z)=-\hf\left[\int_{+\infty}^x\left(z-\m_+ + R(y,z) \right)dy +\m_+ x \right] +K(z)
\ee
for any finite $x\in{\cal B}$ and  $z\in{\cal C}$, where $K(z)$  does not depend on $x$
and the contour of integration is in $\Bscr$. 
Convergence of the integral in \eqref{gxz} follows from \eqref{limzmR}.

Assuming that 
$\a(x)$ and $f_0(z)$ satisfy
 modulation equation \eqref{modeqint}, we want to determine $K(z)$ so that  
$g(x,z)$ defined by \eqref{gxz} coincide with $g(z)$  defined by \eqref{gform}.
Note that 
\be\label{limg1}
\lim_{x\ra +\infty} \left[  g(x,z)+\hf\m_+ x\right]  =K(z)~.
\ee
Rewriting \eqref{gform} as 
 \begin{equation}\label{formg}
g(x,z)=
{{R(x,z)}\over{4\pi i}}
\int_{\gt_m}{{f_0(\z)-x\z}\over{(\z-z)R_+(x,\z)}}d\z~ \ \ \ \ 
\end{equation}
and taking limit of \eqref{formg} as  $ x\ra +\infty$,~~$x\in\R$,
$x\in{\cal B}$,  
we obtain, according to \eqref{limR}, 
\begin{align}\label{limg2}
&\lim_{x\ra +\infty}  \left[  g(x,z)+\hf\m_+ x\right]  = \cr
\lim_{x\ra +\infty}&\left[ \frac{(z-\m_+)(f_0(\m_+)-x\m_+)}{4\pi i(z-\m_+)}\int_{\gt_m}\frac{d\z}{\sqrt{(\z-\a(x))(\z-\overline {\a(x)})}}+\hf\m_+ x\right]
=\hf f_0(\m_+). \cr
\end{align}
Comaring \eqref{limg1} and \eqref{limg2}, we obtain
\be\label{K}
K(z)=\hf f_0(\m_+) \in\R,
\ee
the latter follows from the requirement that $g(x,z)$, $x\in\R$ , is Schwarz-symmetrical in $z$.
Thus,  we obtain
\be\label{gxzfin}
g(x,z)=-\hf\left[\int_{+\infty}^x\left(z-\m_+ + R(y,z) \right)dy +\m_+ x \right] +\hf f_0(\m_+),
\ee
where $f_0(\m_+)$ is a free real parameter.

We want to emphasize that  \eqref{gxzfin} represents a new form of  solution to the RHP 
\eqref{rhpg}-\eqref{modeq} 
(with $t=0$),  written as an integral in the $x$-plane, see Theorem \ref{invertmodeq} below.

Note that for a finite fixed $z\in\bar\C^+$ function $g(x,z)$, defined by \eqref{gxzfin}, is analytic 
for all finite $x\in{\cal B}$ except $x=x(z)$ and $x=\overline{x(\bar z)}$. According to 
\eqref{limR}, we have
\be\label{lim1R}
1+\frac{z-a(x)}{R(x,z)}=\frac{a(x)-\m_\pm}{z-\m_\pm}+o(x^{-p})
\ee
as $\Re x\ra \pm\infty,~x\in{\cal D}$, for any fixed $z\neq \m_\pm$ respectively. Thus,
for any finite $x\in{\cal B}$, function $g(x,z)$, defined by \eqref{gxzfin}, is analytic 
in all finite $z\in {\cal C}$ except $z=\m_+$ and branchpoints $z=\a(x),~z=\tilde \a(x)$.

Now, using $f=2g-h$, we obtain
\be\label{fxz}
f(x,z)=\int^{+\infty}_{x(z)}\left[z-\m_+ + R(y,z) \right]dy +(z-\m_+) x(z)-xz + f_0(\m_+)~,
\ee
so that 
\be\label{finitz}
f_0(z)=\int^{+\infty}_{x(z)}\left[z-\m_+ + R(y,z) \right]dy +(z-\m_+) x(z)+ f_0(\m_+).
\ee
Since $h(x,z)$ and $g(x,z)$ are analytic in $z\in {\cal C}$ except the branchpoints
$z=\a(x),~z=\tilde \a(x)$, the log points and, possibly, points $z=\m_\pm$, function
$f_0(z)= 2g(x,z)-h(x,z)-xz$ is analytic over the same domain. But $f_0(z)$
does not depend on $x$, hence it is analytic everywhere in ${\cal C}$ except
the log points and   possibly, points $z=\m_\pm$, with
\be\label{finitz'}
f'_0(z)=\int^{+\infty}_{x(z)}\left[1 +\frac{z-a(y)}{ R(y,z)} \right]dy +x(z).
\ee


\begin{theorem}\label{invertmodeq}
For any  $\a(x)$ satisfying conditions {\bf A1} - {\bf A4}, functions 
 $g(x,z)$ and $f_0(z)$, given by \eqref{gxzfin} and \eqref{finitz} respectively,
satisfy the RHP \eqref{rhpg}-\eqref{modeq} all $x\in\R$ such that $\a'(x)\ne 0$.
\end{theorem}
 
\begin{figure}
\centerline{
\includegraphics[height=5cm]{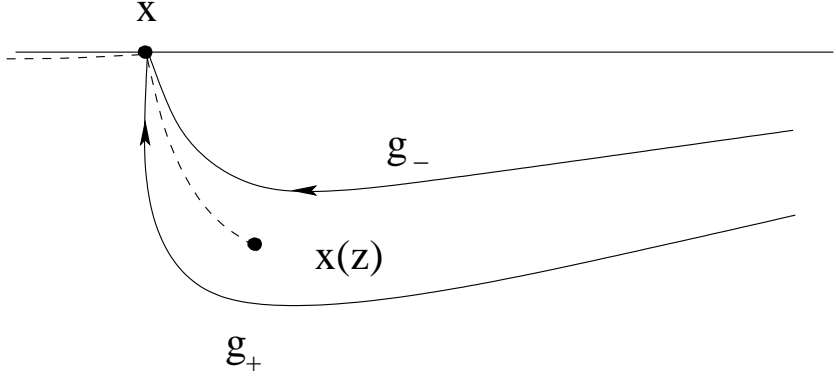}}
\caption{The branchcut of $R$ in ${\cal B}$, connecting $x(z)$ and $-\infty$  is
shown by the dashed line, contours of integration for $g_\pm$ are shown by solid lines.}
\label{gcont}
\end{figure}

\bp
The contour
$\g_m$ in the RHP \eqref{rhpg} can be deformed within the domain of analyticity of 
$f_0$ (endpoints $\a(x), \bar \a(x)$ and the midpoint $\m_+$ remain fixed)
without affecting the solution $g(x,z)$. 
There are no more than a finitely many log points in any compact subset of $\bar{\cal L}$. 
According to the construction, set ${\cal C}$ is connected, i.e., without any loss of
generality we can assume that $\g_m$ lies within the domain of analyticity of $f_0$ (with
the exception of the point $\m_+$). Let us fix some $x\in\R$, such that $\a'(x)\ne 0$, choose some
$z$ on the corresponding $\g_m$, and consider $g_+(z) + g_-(z)$. 
The contours of integration for $g_\pm$ in \eqref{gxzfin} lie on opposite sides of the
branchcut of $R(x,z)$ in the complex $x$-plane ($z$ is fixed), as shown
on Fig. \ref{gcont}. Then
\be\label{gsatisrhpg}
g_+(x,z)+g_-(x,z)=\int^{+\infty}_{x(z)}\left[z-\m_+ + R(y,z) \right]dy+
\int_{x}^{x(z)}(z-\m_+)dy-\m_+x + f_0(\m_+)=f(x,z)~.
\ee
To prove that $g(x,z)$ is analytic at $z=\infty$ for any $x\in{\cal B}\setminus\{x(\infty)\}$,
we fix some $x$ and consider
\be\label{g_z}
\frac{\part}{\part z}g(x,z)=-\hf\int_{+\infty}^x\left( 1+\frac{z-a(y)}{R(y,z)}\right) dy~,
\ee
where the contour of integration does not pass through $x(\infty)$. Then
\be\label{Rzrainf}
R(y,z)=-(z-a(y))+O\left( \frac{b^2(y)}{z-a(y)}\right) 
\ee
as $z\ra\infty$ uniformly on the contour of integration. Therefore, the integrand in
\eqref{g_z} is of the order $O(b^2(y))$ uniformly on the  contour of integration
as $z\ra\infty$, so that $\frac{\part}{\part z}g(x,z)|_{z=\infty}$ is well defined.
Proof of  \eqref{modeq} follows from \eqref{hsamodeq} and the fact that
$f_0(z)$ is analytic  at $z=\a(x)$.
Thus, requirements i) - iii) of the RHP \eqref{rhpg}-\eqref{modeq} are satisfied.
\ep

\begin{remark}\label{rleft}
Theorem \ref{invertmodeq} remains true if in the expressions for $g(x,z)$ and $f_0(z)$
we replace $+\infty$ with $-\infty$ and $\m_+$ with $\m_-$. 
\end{remark}

\begin{remark}\label{invertabel}
For any $z\in{\cal E}^+$, the contour of integration in \eqref{fxz}
is the interval $[x(z),\infty)$ of the real axis. Then
the change of variables  $u=\a(y)$ converts \eqref{fxz}
into the transformation \eqref{xtof}. Note that this is not true for
$z\not\in{\cal E}^+$, since  $\bar u\neq \overline{\a(\bar y)}$ 
for complex $y$.
\end{remark}

\section{Inequalities \eqref{ineq1}-\eqref{ineq2} for  $f_0$}\label{stprf}

In Sections
\ref{AbelHilbert} and 
\ref{inversion} we constructed the semiclassical 
limit of the scattering data  $f_0(z)$
for ZS system \eqref{ZS} with a given potential \eqref{ID}. Let us assume  that potential \eqref{ID}
satisfied conditions {\bf A}. 
In this section we discuss inequalities  \eqref{ineq2} for  $f_0$ as well as existence of
the main and the complementary arcs satisfying \eqref{ineq1} for  all
$x\in\R$.

We start with the observation that for any $x\in\R$ we have
\be\label{imhmu}
\Im h(x,\m_\pm)=-\Im \int_{\pm\infty}^x \sqrt{(\m_\pm-\a(y))(\m_\pm-\overline{\a(y)})}dy=0~,
\ee
since the contour of integration lies on the  real line and the integrand is real valued.  
Convergence of the integral in \eqref{imhmu}
follows from \eqref{ablim}.

Let us consider $w(z), z\in\R$. According to \eqref{ineq2}, \eqref{hxz} and Schwarz symmetry of $g(x,z)$,
$x\in\R$, we have
\be\label{wzalp}
w(z)=\sign(\m_+-z)\Im \int_{x(z)}^x \sqrt{(z-\a(y))(z-\tilde \a(y))}dy,~~
\ee
where $x\in\R$ can be choosen arbitrarily, i.e., the right hand side  of
\eqref{wzalp} does not depend on a particular choice of $x\in\R$. Therefore,
we can choose contour of integration in \eqref{wzalp} as the vertical segment
$[x(z),x]$, where $x=\Re x(z)$, traversed in the negative direction, i.e., down
(note that $\Im x(z)>0$). 

\begin{lemma}  \label{wz>0}
If 
\be\label{arg<pi}
|\arg(z-\a(y))-\arg(z-\a(\bar y))|\le\pi
\ee
for every $y\in[x(z),\Re x(z)]$, where $z>\m_+$, then $w(z)<0$.  If 
\be\label{arg>pi}
|\arg(z-\a(y))-\arg(z-\a(\bar y))-2\pi|\le\pi
\ee
for every $y\in[x(z),\Re x(z)]$, where  $z<\m_-$,  then $w(z)<0$.

\end{lemma}

\begin{figure}
\centerline{
\includegraphics[height=4cm]{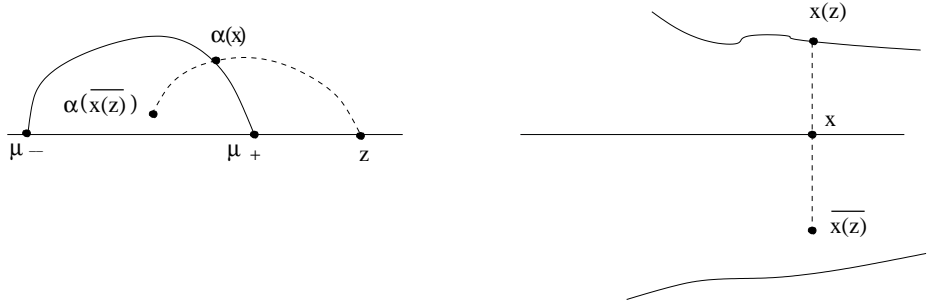}}
\caption{Two solid curves in the $x$-plane show the boundaries of $\Bscr$.
The image of the segment $[x(z),\overline{x(z)}]$ in $z$-plane is shown by
the dashed line.  }
\label{contvert}
\end{figure}

\bp Consider, for example, the case $z>\m_+$. Since $dy$ is negative purely imaginary,
it is sufficient to show that $\Re \sqrt{(z-\a(y))(z-\tilde \a(y))}<0$. Taking into
account our determination of the square root \eqref{brR}, the latter condition is equivalent to
$|\arg(z-\a(y))+\arg(z-\tilde\a(y))|\le\pi$, which, together with \eqref{alphat},
implies \eqref{arg<pi}, see Fig. \ref{contvert}. Similar arguments prove the remaining case $z<\m_-$.
\ep

\begin{remark}\label{suffwz}
Listed below are some sufficient conditions for \eqref{arg<pi} and \eqref{arg>pi} to be true for some  $z>\m_+$ or some $z<\m_-$ respectively.

1. $w(z)<0$ for some  $z>\m_+$ or 
 some  $z<\m_-$ provided 
\be\label{segincl}
[x(z),\overline{x(z)}]\subset\Bscr .
\ee
Indeed, this condition implies that and $\a(\bar y)\in {\cal E^+}$ for any $y\in[x(z),x]$, 
see Fig. \ref{contvert}. Since
$\a(y)\in \C^+\setminus {\cal E}$, 
the inequality \eqref{arg<pi} is satisfied. Similar considerations are valid for \eqref{arg>pi}. Condition \eqref{segincl} is 
satisfied if, for example, the upper boundary $\part\Bscr^+$ satisfies the vertical line test
and if the complex conjugate $\overline{\part\Bscr^+}\subset\Bscr$.

2. According to lemma \ref{wz>0}, $w(z)<0$ for some  $z>\m_+$ 
provided that for every $y\in[x,x(z)]$ the angle between $z-\a(y)$ and 
$z-\a(\bar y)$ is less than $\pi$. This happens, for example, if one can draw a line
$l$ through the point $z$ so that 
the curve $\a(y)$, $y\in[x(z),\overline{x(z)}]$ does not cross $l$ (lies in one the halfplanes
produced by $l$). In particular, if $l$ is a vertical line, we obtain
\be\label{Realphineq}
\Re\a(y)\le z ~~~
\ee
respectively  for all $y\in[x(z),\overline{x(z)}]$.
Similar results hold for $z<\m_-$. 
\end{remark}

We now consider inequality \eqref{ineq1} under additional assumption that 
 $a(x)$ is a monotonically increasing function.
We choose $\g_c^+$
to be the part of $\part{\cal E^+}
\begin{figure}
\centerline{
\includegraphics[height=4cm]{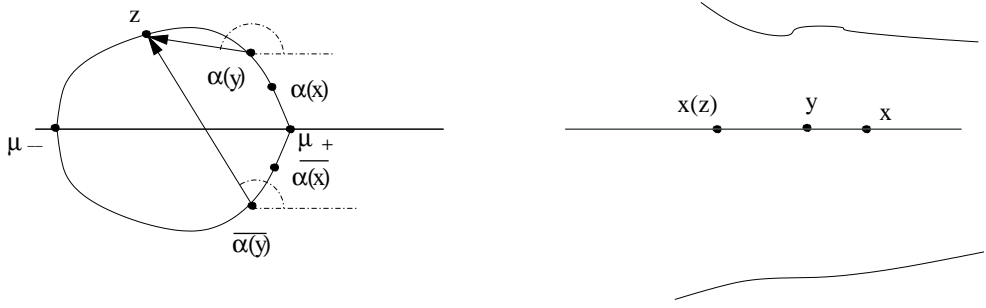}}
\caption{Vectors $z-\a(y)$ and $z-\overline{\a(y)}$ for $y\in[x(z),x]\subset\R$. }
\label{conthor}
\end{figure}

$ connecting $\a(x)$ and $\m_-$.
Take any $z\in\g_c^+$. Then $x(z)$ 
is real and $x(z)<x$. Let us choose any $y\in [x(z),x]$, it is easy to check,
(see Fig. \ref{conthor}) that 
\be\label{argineq}
0\le \arg\left((z-\a(y))(z-\overline{\a(y)})\right)\le 2\pi~,
\ee
so that, taking into account \eqref{hxz} and \eqref{brR}, we obtain  $\Im h(x,z)>0$ on $\g_c$.

Let us now study the sign of $\Im h(x,z)$ on the main arc $\g_m^+$. Notice
that $h_+ + h_-\equiv 0$ on $\g_m$, so that, in general,  $\Im h$ has 
opposite signs on opposite sides of $\g_m$. Therefore, we have to have
$\Im h\equiv 0$ on $\g_m$. Let us first find $\sign \Im h$ on the arc of
$\part{\cal E^+}$ between $\a(x)$ and $\m_+$. Equivalently, we can think of 
deforming $\g_m^+$ to the above mentioned arc of $\part{\cal E^+}$ and finding
$\sign \Im h$ on the negative (external) side of the arc. 
It is easy to see
that if point $z$ will pass over $\a(x)$ outside (of ${\cal E}$), then the
sum of the angles, shown at Fig. \ref{conthor}, i.e.,  $\arg\left((z-\a(y))(z-\overline{\a(y)})\right)$, satisfies inequalities  
\eqref{argineq}. Repeating the previous arguments, we readily obtain 
$\Im h(x,z)<0$, $z\in \part{\cal E^+}$ between $\a(x)$ and $\m_+$ on the 
outer (negative) side of the contour. Of course, on the opposite (positive)
side of the branchcut, $\Im h(x,z)>0$. That means, that inequalities on $\g_m^+$
will be satisfied if there exists a branch of the level curve $\Im h=0$ ($x\in\R$
is fixed), connecting $\m_+$ and $\a(x)$. This question is discussed below.


We now want to calculate $\frac{\part h}{\part z} (x,\m_+^+)$ by taking limit
$z\ra\m_+$ along the negative (outer) side of $\g_m^+$ (it is still asumed that $\g_m$ deformed to coincide with arc of $\part{\cal E^+}$ connecting  $\a(x)$ and $\m_+$).  According to
\eqref{h_z}, 
\be \label{h_z1}
\frac{\part h}{\part z} (x,z)=  \int^{x(z)}_x \frac{z-a(y)} {\sqrt{(z-\a(y))(z-\overline{\a(y)})}}dy ~.
\ee
Therefore, the fact that 
\be\label{h_zarg}
0<\varphi -\hf(\varphi_1+\varphi_2) < \pi~,
\ee
where $\varphi=\arg (z-a(y)),~~\varphi_1=\arg(z-\a(y)),~~\varphi_2=\arg(z-\overline{\a(y)})$,
together with the determination of the proper branch of the radical $R(y,z)$,
imply $ \Im \frac{\part h}{\part z} (x,z)<0$ in \eqref{h_z1}. 

\begin{figure}
\centerline{
\includegraphics[height=6cm]{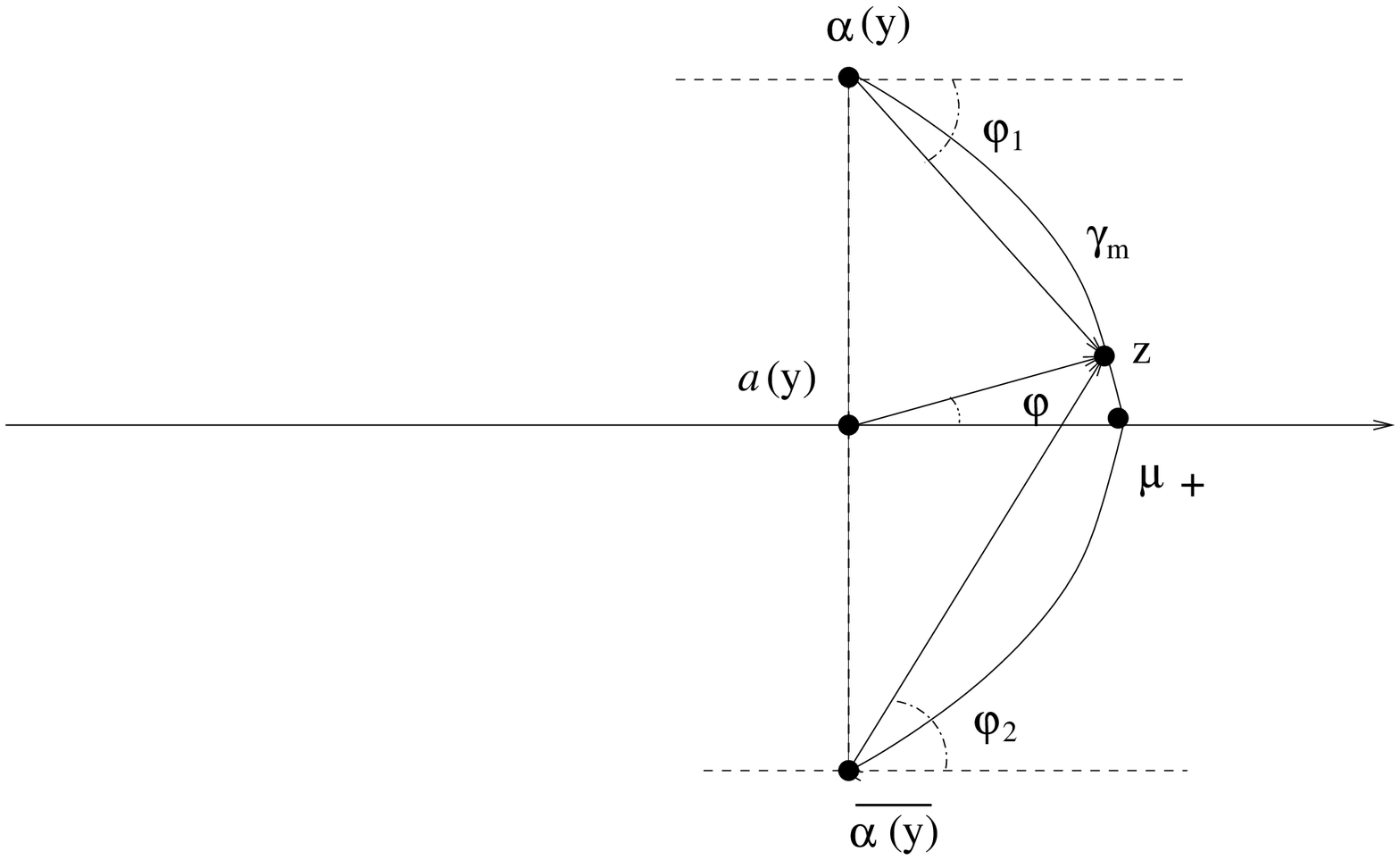}}
\caption{Angles $\varphi_1$ (negative), $\varphi_2$ and $\varphi$ for $\a(y)$, where 
$x\le y\le x(z)$.}
\label{angles}
\end{figure}

\begin{figure}
\centerline{
\includegraphics[height=6cm]{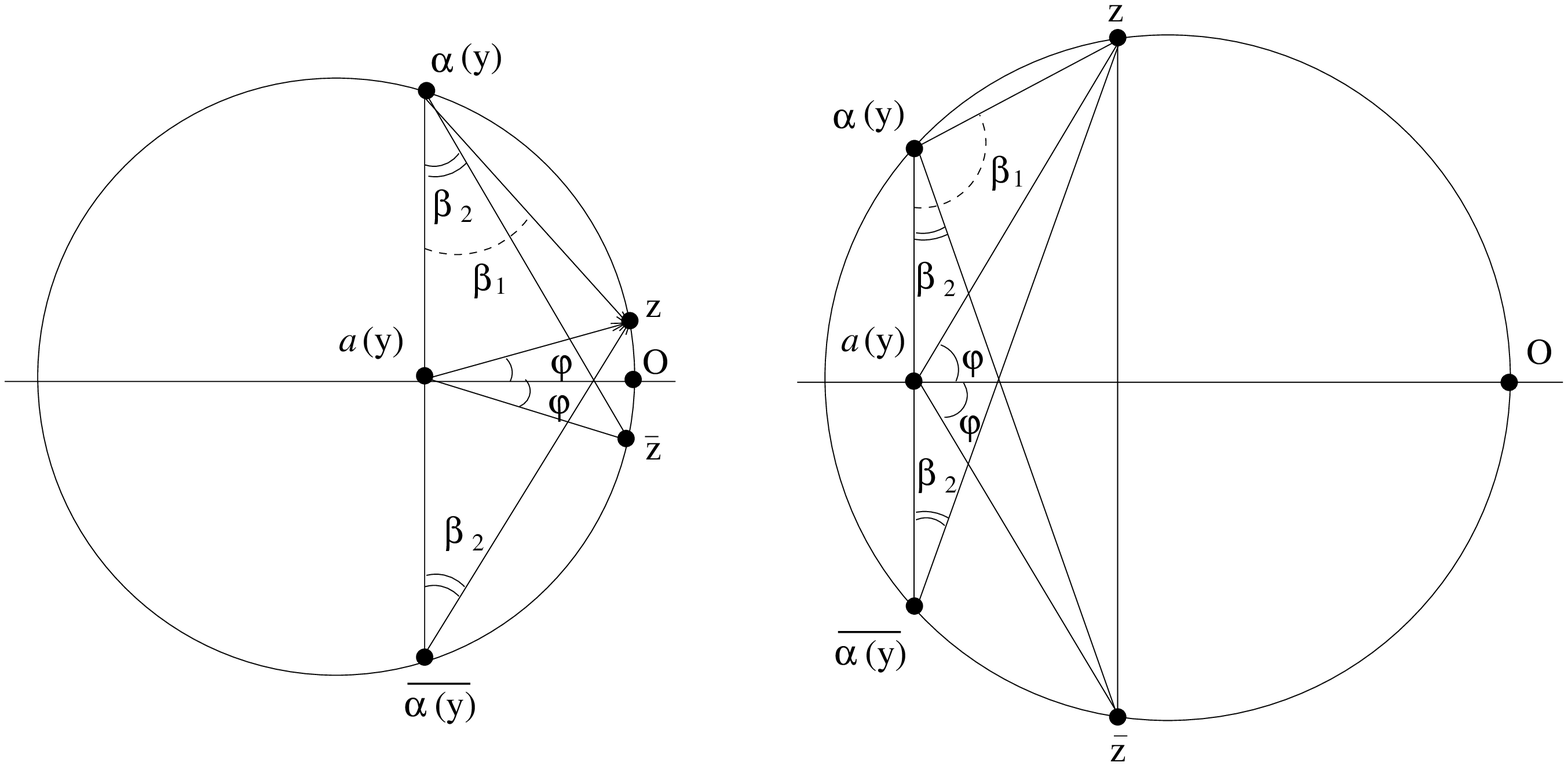}}
\caption{ Triangle $\a(y),z,\overline{\a(y)}$ inscribed in a circle. The left circle shows the case $\varphi_1<0$, the right circle shows the case $\varphi_1>0$.}
\label{inscribed}
\end{figure}

To prove \eqref{h_zarg},  we first
notice that the monotonicity of $a(x)$ on $\R$ implies 
\be\label{ineqphi}
0<\varphi\le\pt,~~-\pt\le\varphi_1<\pt,~~0<\varphi_2\le \pt,~~\varphi_2\ge\varphi\ge|\varphi_1|,	
\ee
see Fig. \ref{angles}, so that the second inequality of \eqref{h_zarg} follows from \eqref{ineqphi}.
Notice that $\b_1>\b_2$, where $\b_1=\pt+\varphi_1$, $\b_2=\pt-\varphi_2$. Then the remaining
 inequality \eqref{h_zarg} becomes 
\be\label{ineqbet}
2\varphi>\b_1-\b_2~.
\ee
Let us inscribe the triangle $\a(y),z, \overline{\a(y)}$ into the circle. As shown on
Fig. \ref{inscribed}, cases $\varphi_1<0$ and $\varphi_1>0$, both angles $\b_1-\b_2$
and $2\varphi$ rest on the arc $z,O,\bar z$. Then \eqref{ineqbet}  follows from the fact
that the vertex of  angle $\b_1-\b_2$ is on the circle, whereas the vertex of  
angle $2\varphi$ is inside the circle.  Thus, \eqref{h_zarg} is proven.

The fact that $ \lim_{z\ra\m_+^+}\Im\frac{\part h}{\part z} (x,z)$, if exists, is negative  implies
\be\label{w'mu++}
w'(\m_+^+)=- \lim_{z\ra\m_+^+}\Im\frac{\part h}{\part z} (x,z)>0`.
\ee
Choosing the other branch of $R(y,z)$, we obtain 
\be\label{w'mu+-}
w'(\m_+^-)=- \lim_{z\ra\m_+^-}\Im\frac{\part h}{\part z} (x,z)<0
\ee
if the limit exists, where $\m_+^-$ is on the positive (inetrior)
part of $\g_m$.

Equations \eqref{imhmu}, \eqref{w'mu++} and \eqref{w'mu+-} show that $\Im h(x,z)$
is negative for real $z$ in a vicinity of $\m_+$. 
Thus, there exists a zero level curve $\l$ of $\Im h(x,z)$ emanating
from $z=\m_+$ into the upper half-plane. Suppose that
$\l\subset\Cscr$ and 
connects $\m_+$ with $\a(x)$. Then the signs of $\Im h$ to the right and to the
left of $\l$ have to be negative, i.e., the first inequality in \eqref{ineq1}
is satisfied.
The results of this section can be summarized by the following statement.

\begin{statement}\label{stat1}
Let  initial data \eqref{ID} be such that: assumptions {\bf (A)} are satisfied
and $S'(x)$ is monotone on $\R$; $w(z)$ satisfies inequalities \eqref{ineq2}. 
If for a given $x\in\R$ a zero level curve $\l$, emanating from $\m_+$,
passes through $\a(x)$ and does not intersect branchcuts of $f_0$, then
$q_0(x,0,\e)$ is   $O(\e)$ approximation of $\tilde q(x,0,\e)$.
\end{statement}

\section{Some geometrical aspects  of transition to a higher genus (breaking) }\label{break}

In this section we consider $t$ assuming nonnegtive values. 
Functions $f$ and $h$ are defined by \eqref{fzxt} and \eqref{hform} respectively.
As it  was established in \cite{TVZ1}, \cite{TVZ3}, transition from the genus zero
to a higher genus (genus 2) occurs when an additional branch of zero level curve 
of $\Im h(x,t,z)$ (the branch that begins and ends at $z=\infty$) intersects with 
the contour $\g^+=\g_m^+\cup\g_c^+$ at some point $z_b=z_b(x,t)$, which is not a log point 
(the case of $g^+$ intersecting a log point requires further investigation).  
 There are at least
four zero level curves of $\Im h(x,t,z)$ passing through the point  
$z=z_b(x,t)$, which implies that $h_z(x,t,z_b)=0$. 
In the case $z_b(x,t)\not =\a(x,t)$, the  point $z_b$ is called
a double point. There are exactly four zero level 
curves  of $\Im h(x,t,z)$ passing through a double point $z_b$ (degenerate cases of multiple level
curves collision are not considered here, however, see \cite{TV3}).
In the remaining case 
 $z_b(x,t)=\a(x,t)$,
a triple point. There are five zero level 
curves  of $\Im h(x,z,t)$ passing through a nondegenerate triple point $z_b(x,t)$. 
The breaking point $x_b,t_b$ on the $x,t$ plane  
that corresponds to a triple point $z_b$ is the starting  point of the breaking curve that separates the genus zero and genus two regions for $t\ge t_b$. Typically, the breaking curve forms a corner at
 $x_b,t_b$. A regular point of the breaking corve corresponds to a double point $z_b$.
The simple modulated wave $q_0(x,t,\e)$ (semiclassical solution), 
given by \eqref{q0}, \eqref{alpAS} fails to approximate
solution $\tilde q(x,t,\e)$ in the genus two region beyond $t= t_b$. In this region $\tilde q$
can be approximated by a  two-phase modulated wave constructed through Riemann theta functions
(see \cite{TVZ1}). The onset of  a  two-phase wave (genus two)  behavior of $\tilde q$ corresponds
to the appearance of a triple point $\a(x_b,t_b)$ on the spectral plane. The main result of this
section is that at the first breaking point $x_b,t_b$ the semiclassical solution $q_0(x,t,\e)$ 
loses its smoothness. More precisely, if $\a(x,t)$ is a triple point for some $x=x_b,t=t_b$
then the derivative $\a_x(x_b,t_b)=\infty$. In other words, genus zero anzatz $q_0(x,t,\e)$
experiences the first break  at $t=t_b$ only if the amplitude of $q_0$ or  derivative of its phase 
(or both) develops an infinite slope at $x=x_b$. The proof uses representation \eqref{hxz} of $h(x,z)=
h(x,t_b,z)$, where $t_b$ is fixed, as an integral in the complex $x$-plane.

\begin{lemma} \label{derh_zalp}
If assumptions {\bf A} for $\a(x)=\a(x,t_b)$ are satisfied then 
for any fixed $x\in\R$ we have 
\be\label{h_zalp}
\frac{\part}{\part z}h(x,z)=\frac{\sqrt{ib(x)}}{\sqrt{2}\a'(x)}\sqrt{z-\a(x)} +O(z-\a(x))^{3/2}
\ee
in a vicinity of the branchpoint $z=\a(x)$, provided $\a(x)$ is not a log point.
\end{lemma}

\bp Since $\a(x)$ is not a log point, we have 
$$x(\a(x)+\d)=x+\frac{dx}{dz}\d + o(\d)=x+\frac{\d}{\a'(x)}+o(\d),$$
where $\d\in\C$ is a small.
Using \eqref{h_z}, we calculate
\begin{align}\label{aaaa}
\left. \frac{h_z(x,z)}{R(x,z)}\right|_{z=\a(x)}=&
\lim_{\d\ra 0}\frac{1}{R(x,\a(x)+\d)}\int_x^{x+\frac{\d}{\a'(x)}}\frac{z-a(y)} {\sqrt{(z-\a(y))(z-\tilde\a(y))}}dy=\cr
&\lim_{\d\ra 0}\frac{ib(x)+\d}{\a'(x)(2ib(x)+\d)}
=\frac{1}{2\a'(x)}~,
\end{align}
which implies \eqref{h_zalp}.
\ep

As an immediate consequence of Lemma \ref{derh_zalp}, we obtain 
\be\label{h(alph)}
h(x,z)=\left[ \frac{\sqrt{2ib(x)}}{3\a'(x)}+O(\sqrt{z-\a(x)})\right] (z-\a(x))^{3/2}~.
\ee

It is well known that zero genus anzatz for zero dispersion limit of the KdV equation
breaks down when it develops infinite slope. The following corollary, which is
another immediate consequence of  Lemma \ref{derh_zalp}, is an  analog of this
statement for the focusing NLS equation.

\begin{corollary}\label{breakcond}
In the conditions of 
 Lemma \ref{derh_zalp} $\a(x_b)=\a(x_b,t_b)$ is  a triple point only if $\a'(x_b)=\infty$. 
\end{corollary}
\bp
$\a(x_b)$ is a  triple point only if the leading term of \eqref{h(alph)} is zero. Since   
$b(x)>0$ on $\R$ (condition {\bf A1}), we conclude  that  $\a'(x_b)=\infty$.
\ep

\begin{remark}
According to \cite{DGC}, solution $q(x,t,\e)$ to Cauchy problem  \eqref{FNLS},
\eqref{ID}  develops {\it elliptic umbilic singularity} at $(x_b,t_b)$ if 
$\a(x_b,t_b)$ is a triple point. It is hypotesized there
that in a vicinity of $(x_b,t_b)$ a solution $q(x,t,\e)$ can be approximated by the special {\it 
tritronqee} solution of the first Painleve equation P1 (see \cite{DGC} for further details).
\end{remark}

\begin{figure}
\centerline{
\includegraphics[height=4cm]{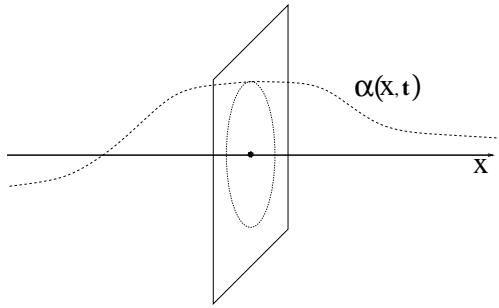}}
\caption{Curve $\a(x,t)$ in $\R\times \C$, spectral plane $\C$ is orthogonal
to real $x$-axis. Genus zero case.}
\label{breakcfg1}
\end{figure}

\begin{figure}
\centerline{
\includegraphics[height=4cm]{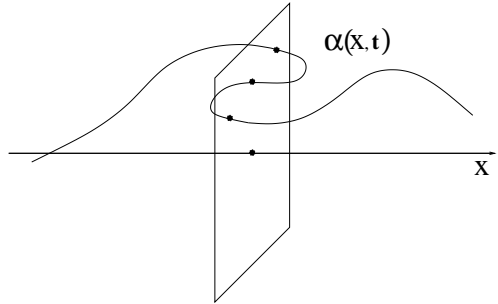}}
\caption{Curve $\a(x,t)$
 in $\R\times \C$, spectral plane $\C$ is orthogonal
to real $x$-axis. The curve developed a ``fold". Three point of intersection
of $\a(x,t)$ and the spectral plane are indicated. For the values of $x$, 
corresponding to the ``fold", points $(x,t)$ belong to the genus two region.}
\label{breakcfg3}
\end{figure}

\begin{figure}
\centerline{
\includegraphics[height=4cm]{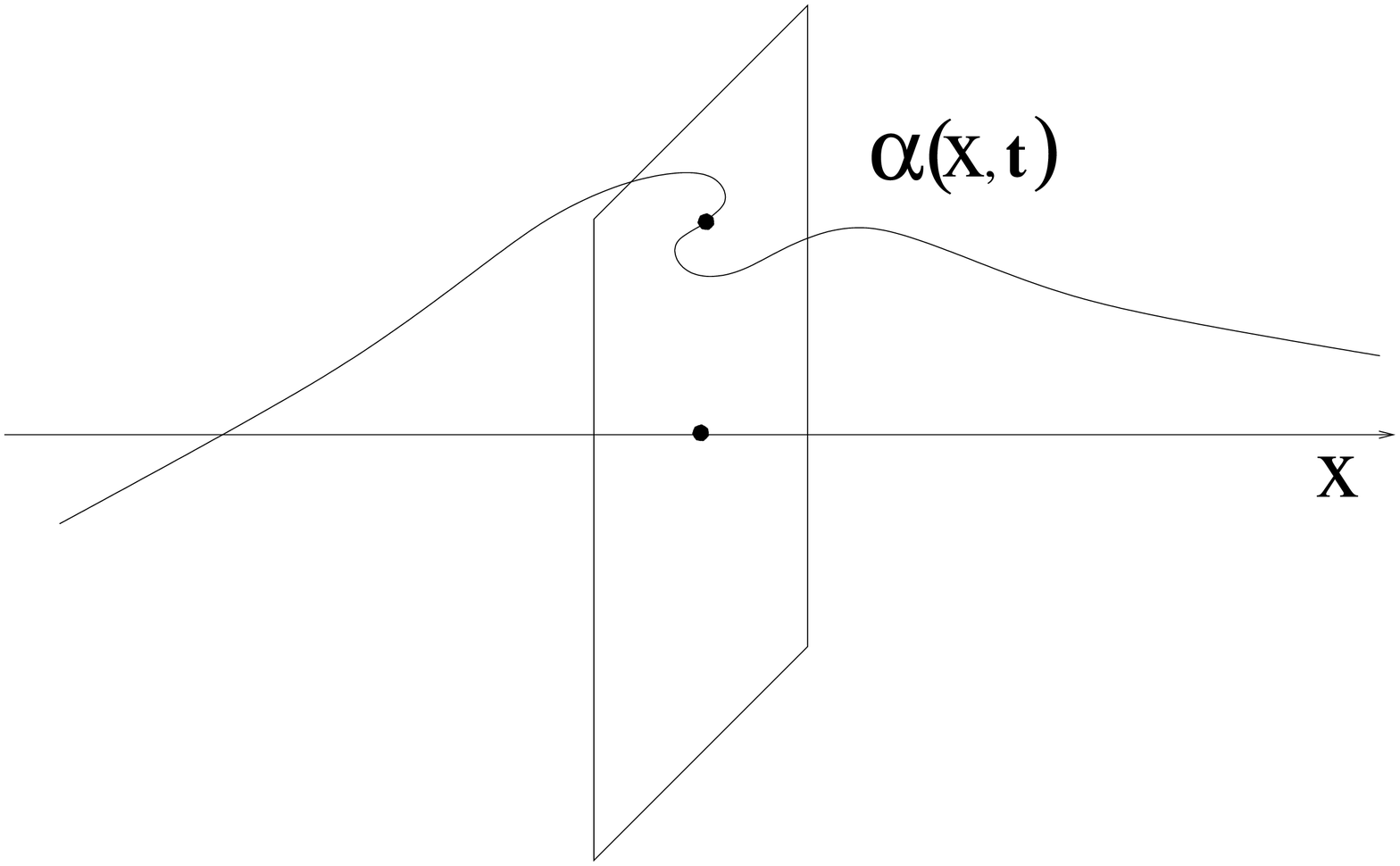}}
\caption{Curve $\a(x,t)$ in $\R\times \C$, spectral plane $\C$ is orthogonal
to real $x$-axis. The point  of tangency between $\a(x,t)$ and $\C$ 
corresponds to a triple point.} 
\label{breakcfg2}
\end{figure}

An obvious geometrical interpretation of Corollary \eqref{breakcond} is the curve
$\a(x)$ in $\R\times \C$, where $x\in\R$ and $\a(x)\in\C$. For any $x\in\R$, this
curve has a unique intersection with the  plane perpendicular to $\R$ at $x$, which 
we interpret as a spectral plane $\C$, see Fig. \ref{breakcfg1}. This property is preserved under the NLS
evolution $\a(x,t)=a(x,t)+ib(x,t)$, 
where $\a(x,0)=\a(x)$, as long as $(x,t)$ is in the genus zero region.  
To get to a higher genus region, the curve $\a(x,t)$ must develop a ``fold" 
(at least in the solitonless case), shown at Fig. \ref{breakcfg3}. It is clear from the
topological point of view that $\a(x,t)$ should become tangential to the spectral
plane at some $(x_b,t_b)$ before the fold can develop, see Fig. \ref{breakcfg2}. The point $(x_b,t_b)$ correspond to a triple point. 
Geometricaly, it is clear that $\a_x(x_0,t_0)=\infty$, which 
is exactly the statement of  Corollary \eqref{breakcond}. 

\section{Symmetry}\label{symmetry}

In this section we establish connections between the symmetry of the initial and the scattering data.

\begin{statement}\label{symalph}
Initial data $q(x,0)=q(x,0,\e)$ given by \eqref{ID} is even iff 
\be\label{alphs}
\a(-\bar x)=-\overline{\a(x)}~.
\ee
for all $x$ in the domain of analyticity $\Dscr$ of $\a(x)$. Here $\a(x)=a(x)+ib(x)=-\hf S'(x)+iA(x)$.
\end{statement}

\bp If $q(x,0)$ is even then $a(x)$ is odd and $b(x)$ is even. Then $\a(-\bar x)= -a(\bar x)+ib(\bar x) =
-[\overline{a(x)}-i\overline{b(x)}]=-\overline{\a(x)}$. Conversly, taking $x\in\R$, we obtain
$a(-x)+ib(-x)=\a(-\bar x)=-\overline{\a(x)}=-a(x)+ib(x)$.\ep

\begin{statement}\label{symx}
Equation \eqref{alphs} is equivalent to 
\be\label{xsym}
x(-\bar z)=-\overline{x(z)}~
\ee
that holds for all $x\in\Bscr$.
\end{statement} 

\bp
$x=x(z)~\Leftrightarrow~z=\a(x)\Leftrightarrow~ -\bar z=-\overline{\a(x)}$. Then, according to 
\eqref{alphs}, $-\bar z=\a(-\bar x)$, so that $x(-\bar z)=-\overline{x(z)}$. Proof of the converse 
statement is similar.\ep

Before proving the next statement, it's worth reminding that the radical $R(x,z)$ with a fixed $x$ has
a branchcut $\g(x)=\g_m$ connecting $\a(x)$ and $\tilde\a(x)=\overline{\a(\bar x)}$ that passes through $\m_+$.

\begin{statement}\label{symR2}
If $\a(x)$ is an odd function or if $\a(x)$ satisfies  \eqref{alphs} then
\be\label{R2sym}
R^2(-\bar x,-\bar z)=\overline{R^2(x,z)}~,
\ee
for all $\Im z\ge 0$ and all $x\in\Dscr$. The converse is also correct. 
\end{statement} 

\bp Equation \eqref{R2sym} can be written as
\be\label{quadrpoly}
[(z-\a(x))(z-\overline{\a(\bar x)})]=\overline{[(\bar z +\a(-\bar x))(\bar z+\overline{\a(-x)})]}~.
\ee
The proof follows from equating coefficients of linear and free terms (in $z$). \ep

\begin{remark}\label{notodd}
It is easy to verify using the moment conditions for $\a(x)$ from \cite{TVZ3} that odd $\a(x)$
would imply $w(0)=0$ and  
\be
\int_{-\infty}^{\infty}\frac{w'(\z)}{|\z|}d\z=0~.
\ee
\end{remark}

Equation \eqref{R2sym} obviously implies
\be\label{Rsym}
R(-\bar x,-\bar z)=\pm\overline{R(x,z)}~.
\ee
Based on our choice of the branchcut of $R$, it is clear that in the case 
of  real $x,z$ the sign in \eqref{Rsym} is positive if $|z|<\m_+$ and negative if $|z|>\m_+$. 
In the case $x\in R$ the branchcut  $\g(x)$ is symmetrical with respect to $\R$
and intesect $\R$ only at $z=\m_+$.
If $x\in\R$ and $\Im z>0$, the sign in \eqref{Rsym} is positive only if 
there exists a curve $\s(z)$ connecting $z$ with the origin, such that $\s(z)\cap \g(x)=
\emptyset$ and simultaneously $-\overline{\s(z)}\cap \g(-\bar x)=\emptyset$. It is clear
that if such $\s(z)$ does not exist then $\s(z)$ can be choosen in such a way that   $\s(z)\cap \g(x)=
\emptyset$ and $-\overline{\s(z)}$ intersects  $\g(-\bar x)$ only one time. Then the sign in
\eqref{Rsym} is negative. In the case of a complex $x\in\Bscr$ the sign in \eqref{Rsym} is defined as above.

Let $\S_*(x)$ denote the region (not necessarily simple) 
bounded by the curve $\g(-\bar x)\cup \{-\overline {\g(x)}\}$, and let 
$\S(x)=(\S_*(x)\cup \{-\overline {\S_*(x)}\}\cap \bar\C^+$. Clearly, 
$\S(x)$ is symmetrical with respect to imaginary axis and has the segment $[-\m_+,\m_+]$
a its lower boundary.
Then, for every $x\in\Bscr$, equation \eqref{R2sym} has positive sign if $z\in \S(x)$
and negative sign if $z\not\in \S(x)$. 

\begin{theorem}\label{symh}
If $\a(-\bar x)=-\overline{\a(x)}$ then 
\be\label{hsym}
h(-\bar x,-\bar z)=\mp \overline{h(x,z)}~,
\ee
where the sign is negative if $z\in \S(x)$ and positive otherwise.
Conversly, \eqref{hsym} implies that $\a(x)$ is odd or $\a(-\bar x)=-\overline{\a(x)}$.
\end{theorem}

\bp
According to \eqref{hxz} and to Statements \ref{symx}, \ref{symR2},
\be\label{alptoh}
h(-\bar x,-\bar z)=\int_{-\bar x}^{-\overline{x(z)}}R(u,-\bar z)du=
-\int_x^{x(z)}R(-\bar y,-\bar z)d\bar y=\mp\int_x^{x(z)}\overline{R(y,z)dy}=
\mp\overline{h(x,z)}~,
\ee
where $y=-\bar u$. The last to terms of \eqref{alptoh} have sign minus if $z\in \S(x)$
and sign plus if $z\not\in \S(x)$.

To prove the converse, we need to show that \eqref{alptoh} implies \eqref{R2sym}.
Let us choose some $x=iy,~y\in\R$,  $z\in \S(x)$ and let us introduce
\be\label{mintro}
m(y,z)=-ih(iy,z)=-ih(x,z)~.
\ee
Then 
\be
h(-\bar x,-\bar z)=h(i\bar y,-\bar z)=im(\bar y,-\bar z)~~~
 -\overline{h(x,z)}=i\overline{m(y,z)}~,
\ee
so, according to \eqref{alptoh}, 
\be\label{msym}
m(\bar y,-\bar z)=\overline{m(y,z)}~.
\ee
Since $m$ is analytic in $y$, we have Taylor expansion $m(y,z)=\sum_{k=0}^\infty a_k(z)y^k$,
so that \eqref{msym} is equivalent to
\be\label{asym}
\overline{a_k(z)}=a_k(-\bar z)~~~~~~~~~~~~~~~~~ \forall k\in \N~.
\ee
Since $m_y(y,z)=\sum_{k=1}^\infty ka_k(z)y^{k-1}$ has Taylor coefficients satisfying 
\eqref{asym}, we obtain 
\be \label{mysym}
m_y(\bar y,-\bar z)=\overline{m_y(y,z)}~.
\ee
But $m_y(y,z)=h_x(iy,z)=h_x(x,z)=-R(x,z)$. Thus, \eqref{mysym} implies \eqref{Rsym} with 
positive sign. By analyticity of $h$, this result can be extended from purely imaginary
to all $x$. Similarly, we can obtain \eqref{Rsym} with 
negative sign when $z\not\in \S(x)$. Thus, \eqref{hsym} implies \eqref{R2sym}.
Statement \ref{symR2} completes the proof. \ep

\begin{corollary}\label{symw}
In the case $x,z\in\R$ equation \eqref{hsym} implies that $w(z)=\sign(z-\m_+)\Im h(x,z)$
is an even function and 
\begin{align}\label{wsym}
\Re h(-x,-z)=-\Re h(x,z)  &~~~~{\rm if}~~~|z|<\m_+\cr
\Re h(-x,-z)=\Re h(x,z)  &~~~~{\rm if}~~~|z|>\m_+~.
\end{align}
\end{corollary}

Corollary \ref{symw} states that $\Im h(x,z)$ is even on 
$z\in[-\m_+,\m_+]$ and is odd outside  
$[-\m_+,\m_+]$. Therefore, $h_z(x,z)$ is odd on $z\in[-\m_+,\m_+]$ and is even outside
$[-\m_+,\m_+]$. 
Taking into account Remark \ref{notodd} and Statement \ref{symalph}, we obtain the following 
corollary.

\begin{corollary}\label{qevenpre}
If $\Im h(0,0)\neq 0$ then \eqref{hsym} implies that the initial potential $q(x,0)$
is an even function.
\end{corollary}

Given a scattering data (see \cite{TVZ3}), we know $w(z)$ but not necessarily $h(x,z)$.
So, does even $w(z)$ implies even $q(x,0)$ under the inverse scattering procedure of \cite{TVZ1}, \cite{TVZ3}?
According to \eqref{hxz}, $h(x,z)$ depends on the branchcut of $R(x,z)$. Let us denote by
$R_R,R_L$ and $h_R,h_L$ the radical $R(x,z)$ (and  corresponding to it $h(x,z)$)
with the branchcut passing through $\pm\m_+$ respectively (in the case of even $w(z)$ we have $\m_-=-\m_+$). 
In all the statements above, we 
considered $R=R_R$ and $h=h_R$. In the case of $x,z\in\R$ we choose $R_L(x,z)$ to be positive
for $z>-\m_+$ and preserve the orientation of contour $\g^+$ (from $\m_+$ to $-\m_+$)
in the matrix RHP for the inverse scattering transform. With such choice of $R_L$,
it follows from \eqref{hxz} that $h_L(x,z)=\pm h_R(x,z)$ for $x,z\in\R$ and $|z|<\m_+$ or 
$|z|>\m_+$ respectively.
If $\a(x)$ satisfies \eqref{alphs}, then, according to Theorem \ref{symh},
 $h_L(-\bar x, -\bar z)=h_R(-\bar x, -\bar z)= -\overline{h_R(x,z)}$ for $z\in \S(x)$ and
$h_L(-\bar x, -\bar z)=-h_R(-\bar x, -\bar z)= -\overline{h_R(x,z)}$ for $z\not\in \S(x)$.
For $x\in\R$ that means 
\be\label{hRL}
\Im h_L(-x,-\bar z)=\Im h_R(x,z)
\ee
for all $z$. Thus, we can formulate a stronger version of
Corollary \ref{qevenpre}.

\begin{corollary}\label{qeven}
If $w(z)$, $z\in \R$, is even and if for all $x\ge 0$ the inverse 
scattering procedure of \cite{TVZ1}, \cite{TVZ3} produces a semiclassical solution  \eqref{ID}
with $a(x)=-\hf S'(x)$ and $b(x)=A(x)$, such that $\a(x)$ is analytic in a region containg $x\ge 0$
 and satisfies \eqref{alphs} in a vicinity of $x=0$, then $q(x,0)$, evenly continued to the whole $\R$,
is the  initial potential corresponding to the scattering data $w(z)$ 
according to the procedure of \cite{TVZ1}, \cite{TVZ3}.
\end{corollary}

\bp Let $\a(x)$,  $x\ge 0$, be solution of the modulation equations (moment conditions),
expressed in terms of $w(z)$, see \cite{TVZ3}, Sect. 3.1.1. 
By Remark 3.1 of \cite{TVZ3},
$\a(x)=-\overline{\a(-x)}$ satisfies modulation equations when $x<0$. Thus, we can assume 
that $\a(x)$, $x\in\R$, is analytic on $\R$ and satisfies \eqref{alphs}. 
According to Theorem \ref{symh}, the corresponding $h_R$ and $h_L$ satisfy \eqref{hsym}.
By the assumption, the main and complementary arcs have the required distribution of signs
of $\Im h_R(x,z)$ for all $x\ge 0$.
Then, according to \eqref{hRL}, in the case of $x<0$ the main and complementary arcs have 
the required distribution of signs
of $\Im h_L(x,z)$. 
Thus, the initial data represented by $\a(x)$, $x\in\R$, indeed corresponds to the given
scattering data $w(z)$ through the inverse scattering procedure of \cite{TVZ1}, \cite{TVZ3}.
\ep

\section{Examples}\label{examples}

\subsection{Example of \cite{TVZ1}, $\m=2$}\label{exam1}

The case of 
\be\label{ourdata}
a(x)=\mt \tanh x,~b(x)=\sech(x),
\ee
$\m>0$,  was studied in details in \cite{TVZ1}.
The leading order (in $\e$) 
\begin{equation}
\label{eq1}
f_0 (z)=\lim_{\e\ra 0}\frac{i\e}{ 2}\ln r_{init} (z,\e)~,
\end{equation}
where $r_{init}(z,\e)$ is the  reflection coefficient of Zakharov-Shabat problem
for the focusing NLS with the inital data \eqref{ID}, $A(x)=b(x)$ and $-\hf S'(x)=a(x)$,
is calculated there to be
\begin{align}\label{f1}
f_{0} (z)=&(\mt-z)\left[\ipt+\ln(\mt-z)\right]
+\frac{z+T}{2}\ln(z+T)+\frac{z-T}{2}\ln(z-T) \cr
&-T\tanh^{-1}\frac{T}{\mt}
+\mt\ln 2+\pt\e, ~~~~~~~{\rm when} \ \Im z\ge 0,  \cr
\end{align}
where $T=\smf$. This calculation involved exact solution of Zakharov-Shabat problem
in terms of hypergeometric functions, followed by Stirling's asymptotic formula.
Note that $f_0(z)$ has a log point $T\in{\cal E}$ in the case $\m< 2$  and has no log
points in ${\cal E}$ in the case $\m\ge 2$. This is the  reflection of the fact that the initial data
\eqref{ourdata} is purely radiative in the case $\m\ge 2$, and contains points of
the descrete spectrum (solitons) on the vertical segment $[-T,T]$ 
in the case $\m< 2$ (see \cite{TV}).
From \eqref{f1} one readily obtain
\begin{equation}\label{f'ex}
f_0 '(z)=-\ipt-\ln(\mt-z)+\hf\ln(z^2-T^2)~ ,
\end{equation}
and 
\be\label{w1'}
w'(z)=-\pt\sign z\left( 1-\chi_{[-T,T](z)}\right),~z\in\R. 
\ee
Since the asymptotic inverse scattering transform, developed in \cite{TVZ1}, \cite{TVZ3},
uses only the values $w'(z),~z\in\R$, and, in the case $\m<2$, the jump of $f_0'(z)$  
over the slit $[0,T]$, we will focus on calculating these quantitees.

Consider first the simplest case $\m=2$. Then $\m_\pm=\pm 1$,
\be\label{alpm=2}
\a(x)=\frac{\sinh x +i}{\cosh x},~~~~\tilde\a(x)=\frac{\sinh x -i}{\cosh x}
\ee
and 
\be\label{alp'm=2}
\a'(x)=\frac{1-i\sinh x}{\cosh^2 x}~.
\ee

\begin{figure}
\centerline{
\includegraphics[height=4cm]{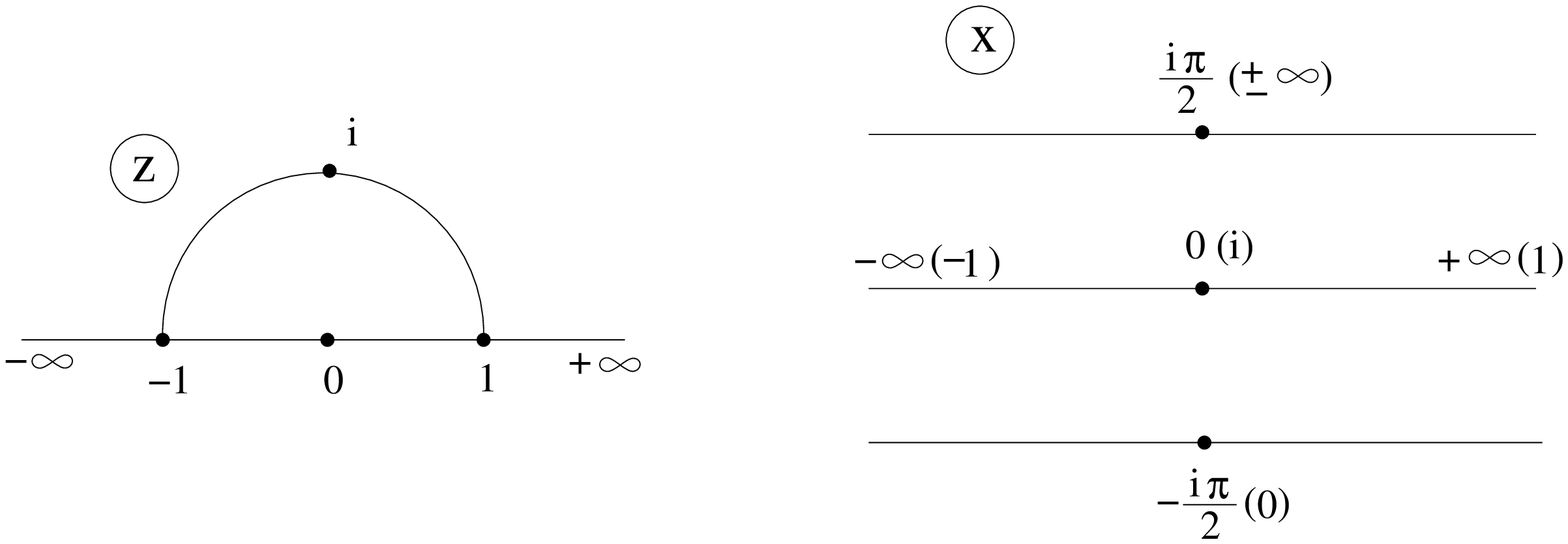}}
\caption{The strip $\Bscr=\{x:~ -\pt\le \Im x\le \pt \}$ is mapped onto the upper
half plane of $z$-plane by $z=\a(x)$, so that $\a(\Bscr^-)={\cal E^+}$. 
Shown in parentheses 
are the  corresponding images of the map $\a(x)$, i.e.,
the preimages of the inverse to $\a(x)$ map $x(z)$.} 
\label{Bscrm2}
\end{figure}

According to \eqref{alpm=2}, $\a(x)$ is a meromorphic function with poles
at $\ipt+2\pi im$ and zeroes at  $-\ipt+2\pi im$, $m\in\Z$. 
Since  
\be\label{alpm=2dB}
\a(\x\pm\ipt)=\frac{\cosh \x\pm 1}{\sinh \x}~,
\ee
we see that $\Bscr$ is the strip $-\pt\le \Im x\le \pt$, where $\ipt=x(\pm\infty)$,
$-\ipt=x(0)$ and $\pm\infty=x(\pm 1)$ respectively, see Fig \ref{Bscrm2}. Since $\a'(x)\neq 0$ inside $\Bscr$, there are no branchpoints in $\Bscr$. Note that
for all $x\in\R$ we have 
\be\label{Em2}
a^2(x)+b^2(x)=1~,
\ee
so that ${\cal E}$ is the unit disc.

In the case $\m=2$, the inverse function $x(z)$ to $\a(x)$ on $\C^+$ is very
simple.  In order to find it,  we calculate
\be\label{abeqm2}
(z-\a)(z-\tilde\a)=(z-a)^2+b^2=z^2-2az+1=0~,
\ee
which yields solution $a(z)=\hf[z+\frac{1}{z}]$. (Here we use notation  $a(z)$ for   $a(z)=
a(x(z))$.)
Thus
\be\label{xzm2} 
x(z)=\tanh^{-1}a(z)=\tanh^{-1}\hf[z+\frac{1}{z}]~.
\ee

\begin{figure}
\centerline{
\includegraphics[height=1cm]{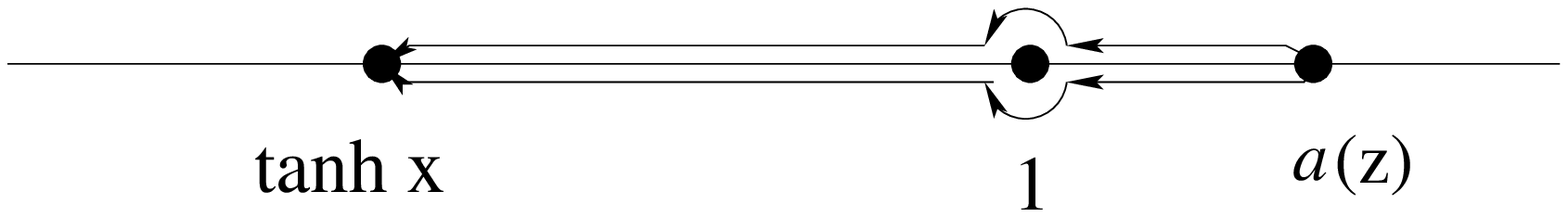}}
\caption{Upper contour of integration is  for $z>1$, lower -  for $z<1$.} 
\label{res1}
\end{figure}

According to \eqref{wzalp}, we have 
\be \label{wzm2}
w(z)=\sign(1-z)\Im \int_{x(z)}^x\sqrt{z^2-2z\tanh y+1 }dy=
\sign(1-z)\Im \int_{a(z)}^{\tanh x}\sqrt{z^2-2z\eta+1 }\frac{d\eta}{1-\eta^2}~,
\ee
where $z\in\R$, $x\in\R$ is arbitrary and $y=\tanh^{-1}\eta$. 
Let us consider the case $z>0$. Then $a(z)>1$, so the contour of integration 
in the latter integral is along the real segment $[\tanh x, a(z)]$ of the complex
$\eta$-plane (complex $a$-plane), except the singular point $\eta=1$, which is traversed from above or
from below if $z>1$ or $z<1$ respectively, see Fig. \ref{res1}. Since the integrand is real on $\R$,
only integration around $\eta=1$ contribute to the imaginary part of the integral.
Thus
\be \label{wzm2+}
w(z)=\pi i\left. \Res\frac{\sqrt{z^2-2z\eta+1}}
{1+\eta}\right|_{\eta=1}=\ipt(1-z)~,
\ee
where the square root is choosen positive when $z<1$ and negative when $z>1$ respectively. By similar argument (but without  $\sign(1+z)$ involved), we obtain
\be \label{wzm2-}
w(z)=\ipt(1+z)~
\ee
for $z<0$.
Thus, $w(z)$ obtained through \eqref{wzalp} with initial data from \eqref{abeqm2} coincides with  $\Im f_0(z),~z\in\R$, where $f_0$ is given by \eqref{f1} with 
$\m=2$ (and $T=0$). The correct distribution of signs of $\Im h(x,z)$  along the main 
and complementary arcs (inequality \eqref{ineq1}) follows directly from the analysis of zero level 
curves of $\Im h(x,z)$ with a fixed $x\in\R$ in the upper $z$-halfplane (three zero level curves
entering $\C^+$ at $\pm 1$ and $\infty$ respectively meet at the branchpoint $\a(x)$).
One can also observe that $f_0(z)$ given by \eqref{f1} can be obtained through the Cauchy 
transform of $w(z)$  (see \cite{TVZ3}), and that sign distribution \eqref{ineq1} for this
$f_0(z)$ was proven in \cite{TVZ1}.

\subsection{Example of \cite{TVZ1},  general case}\label{exam1gen}

We consider now  the general case $\m>0$ in \eqref{ourdata}. Then \eqref{alpm=2} -
\eqref{abeqm2} become 
\begin{align}\label{eqsmugen}
\a(x)=\frac{\mt\sinh x +i}{\cosh x},~~~~&~~~~\tilde\a(x)=\frac{\mt\sinh x -i}{\cosh x},\cr
\a'(x)=\frac{\mt-i\sinh x}{\cosh^2 x},~~~~&~~~~\a(\x\pm\ipt)=\frac{\mt\cosh \x\pm 1}{\sinh \x},\cr
\frac{4}{\m^2}a^2(x)&+b^2(x)=1,\cr
(z-\a)(z-\tilde\a)=(z-a)^2+b^2=&z^2-\m z\tanh x+1+T^2\tanh^2x=0~\cr
\end{align}
respectively. The latter equation implies 
\begin{align}\label{thxz}
\tanh x(z)=\frac{\mt z\pm\sqrt{z^2+\left(1-\msq\right) }}{\msq-1}~,
\end{align}
where the correct branch of the radical in the right hand side should be chosen.
Equations \eqref{eqsmugen} show that ${\cal E}$ is the ellipse centered at $z=0$
with horizontal and vertical semiaxes $\mt$ and $1$ respectively. The complement of 
${\cal E^+}$ in $\C^+$ is mapped onto the strip $0\le \Im x \le \pt$ in the $x$-plane.

To find the image of ${\cal E^+}$, we need to study the branchpoints $x^*$ 
of $x(z)$ defined by 
\be\label{bpointsmu}
\sinh x^*=-\mt i~.
\ee
The images of branchpoints in $z$-plane (log points), obtained by substituting \eqref{bpointsmu} into the expression for $\a(x)$, are
\be\label{logpointsmu}
 z^*=\pm \sqrt{\msq-1}=\pm T~.
\ee

\begin{figure}
\centerline{
\includegraphics[height=5cm]{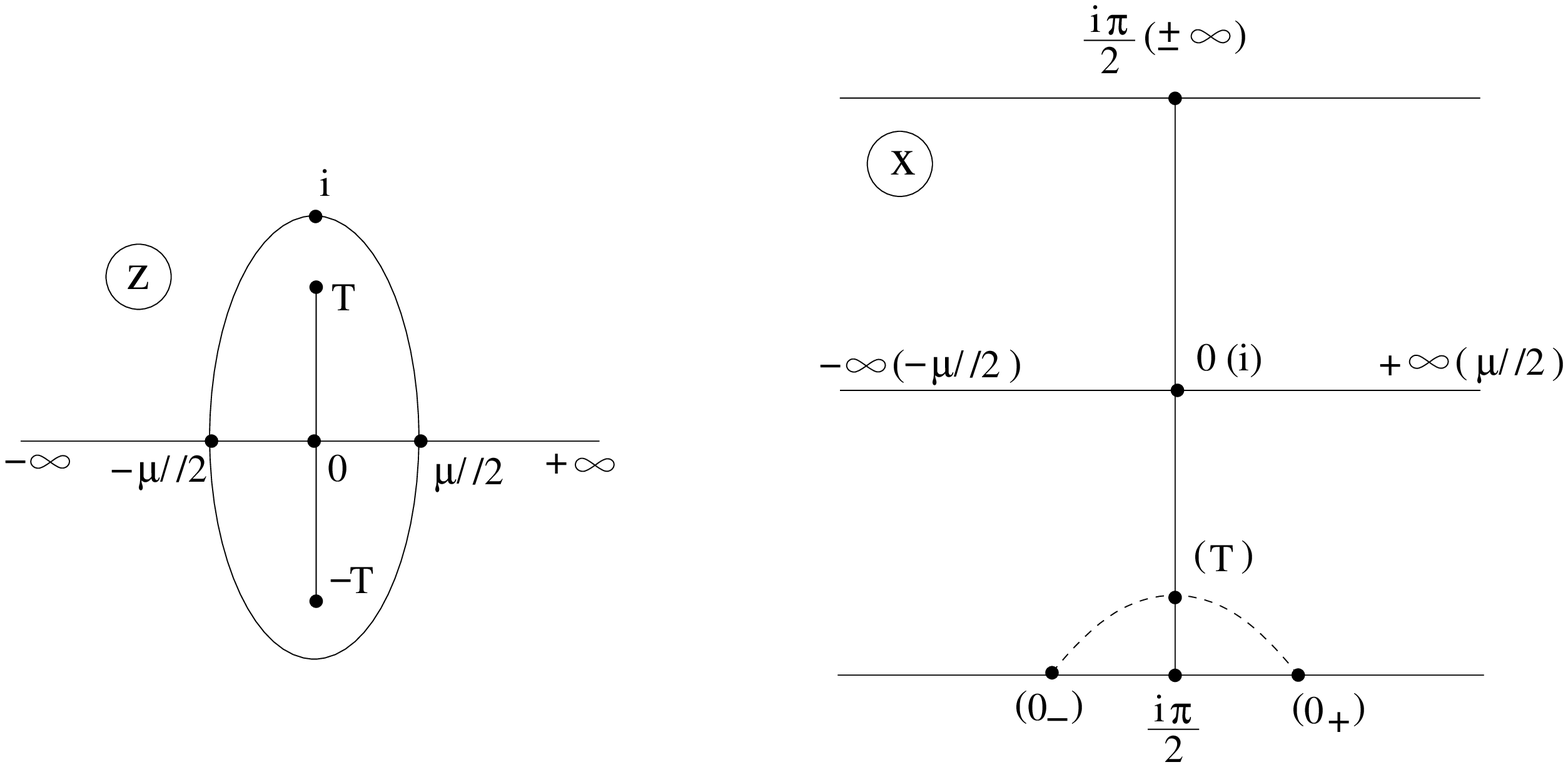}}
\caption{Mapping $\a(x)$, case $\m<2$. Shown in parentheses 
are the  corresponding images of the map $\a(x)$, i.e.,
the preimages of the inverse to $\a(x)$ map $x(z)$.
Set $\Bscr$ is the strip $\{x:~ -\pt\le \Im x\le \pt \}$
without the region bounded by the dashed contour, which is the 
image of two sides of the slit $[0,T]$ by the map $x(z)$.} 
\label{Bscr<}
\end{figure}

\begin{figure}
\centerline{
\includegraphics[height=5cm]{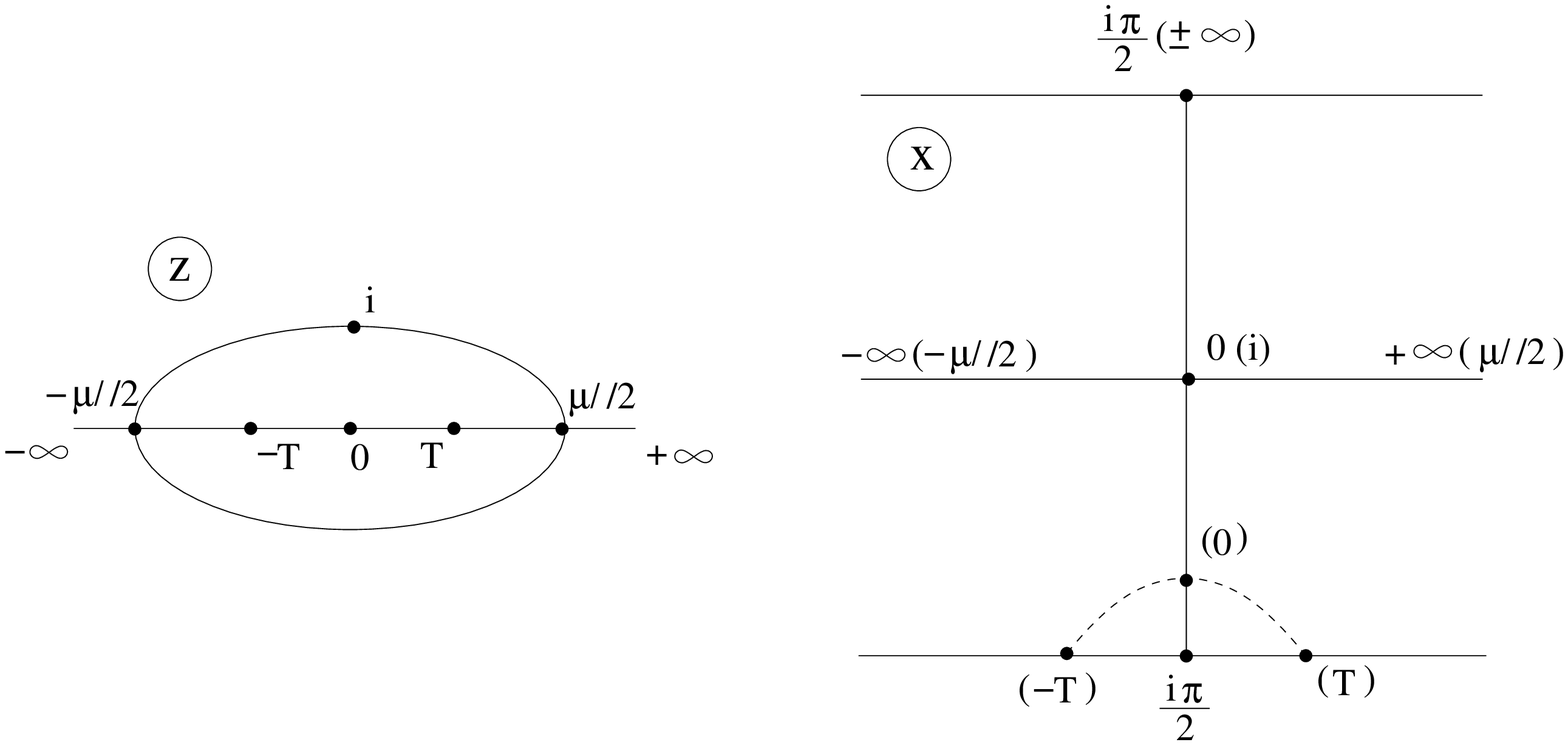}}
\caption{Mapping $\a(x)$, case $\m>2$. Shown in parentheses 
are the  corresponding images of the map $\a(x)$, i.e.,
the preimages of the inverse to $\a(x)$ map $x(z)$.
Set $\Bscr$ is the strip $\{x:~ -\pt\le \Im x\le \pt \}$
without the region bounded by the dashed contour, which is the 
image of the segment $[-T,T]$ by the map $x(z)$. }
\label{Bscr>}
\end{figure}

In the case $\m>2$ log points $\pm T\in(-\mt,\mt)$ are not in the upper halfplane. In the case $\m<2$ log points
$ \pm T=\pm i|T|$. In the latter case, we make a branchcut of $x(z)$
in $\C^+$ over the segment $[0,T]$.
It is now easy to find the image $\Bscr^-$ of the region ${\cal E^+}$ (with the
cut if $\m<2$), see  Fig. \ref{Bscr<}, Fig. \ref{Bscr>} for $\m<2,~\m>2$
respectively. Equations of the arc, connecting
points $x(0^\pm)$ and $x(T)$, case $\m<2$, and $x(\pm T)$ and $x(0)$, case $\m>2$,
are obtained from conditions $\Re \a(y)=0$ and $\Im \a(y)=0$ respectively.
These equations are 
\be\label{brnchcteq}
-\sin \eta=\mt\cosh \x~~~~{\rm  and}~~~~ \cosh\x=\mt\sin\eta
\ee
respectively, where $y=\x+i\eta$.  

\begin{figure}
\centerline{
\includegraphics[height=5cm]{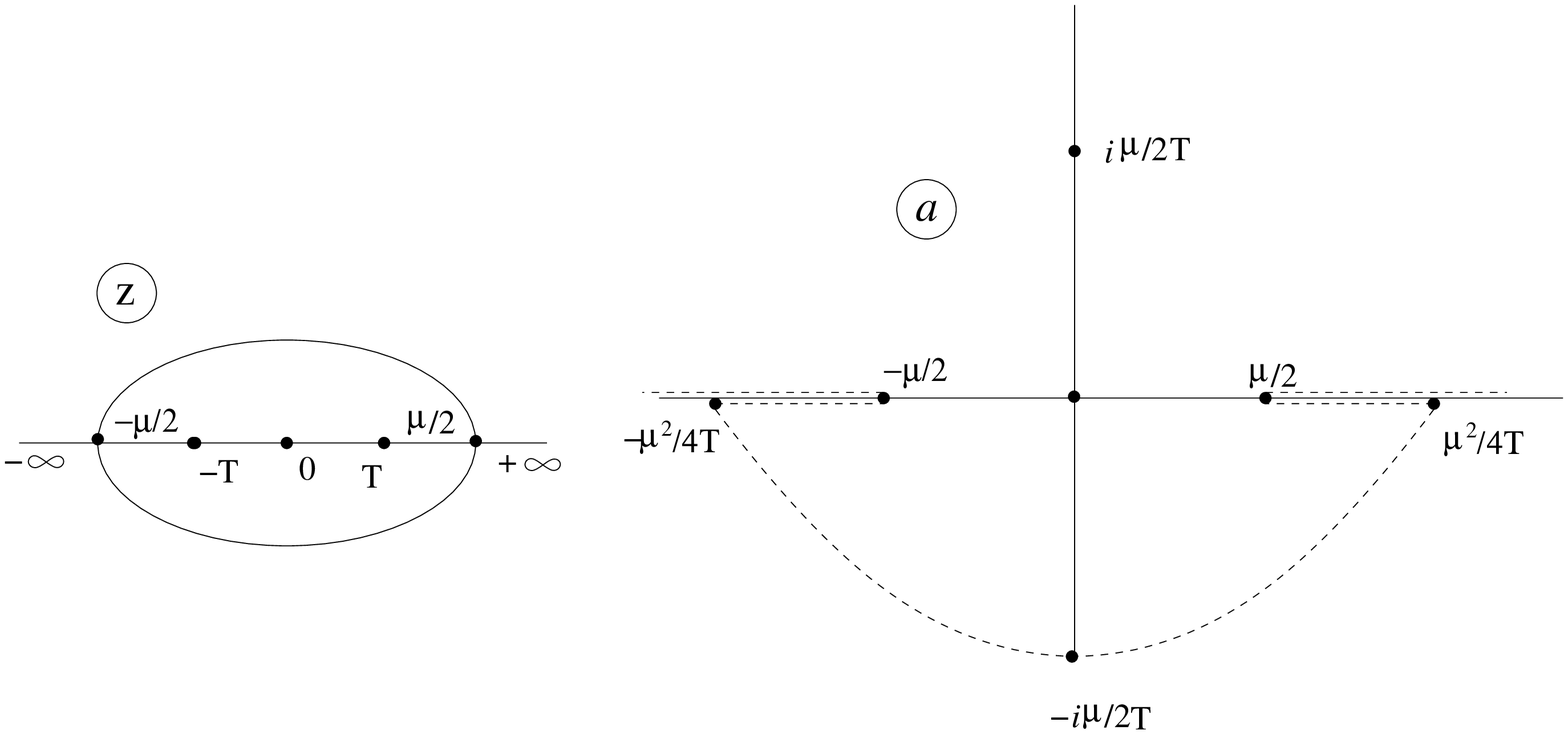}}
\caption{Mappings \eqref{mapaz} in the case $\m>2$. The dashed line
in the $a$-plane is the image of $\R$ by the map $\hat a(z)=a(x(z))$.
In particular,
$\hat a(\pm\infty)=\pm\infty,~~\hat a(\pm\mt)=\pm\mt,~~\hat a(\pm T)=\pm\frac{\m^2}{4T},
~~\hat a(0)=-\frac{i\m}{2T}$.} 
\label{mapam>}
\end{figure}

\begin{figure}
\centerline{
\includegraphics[height=5cm]{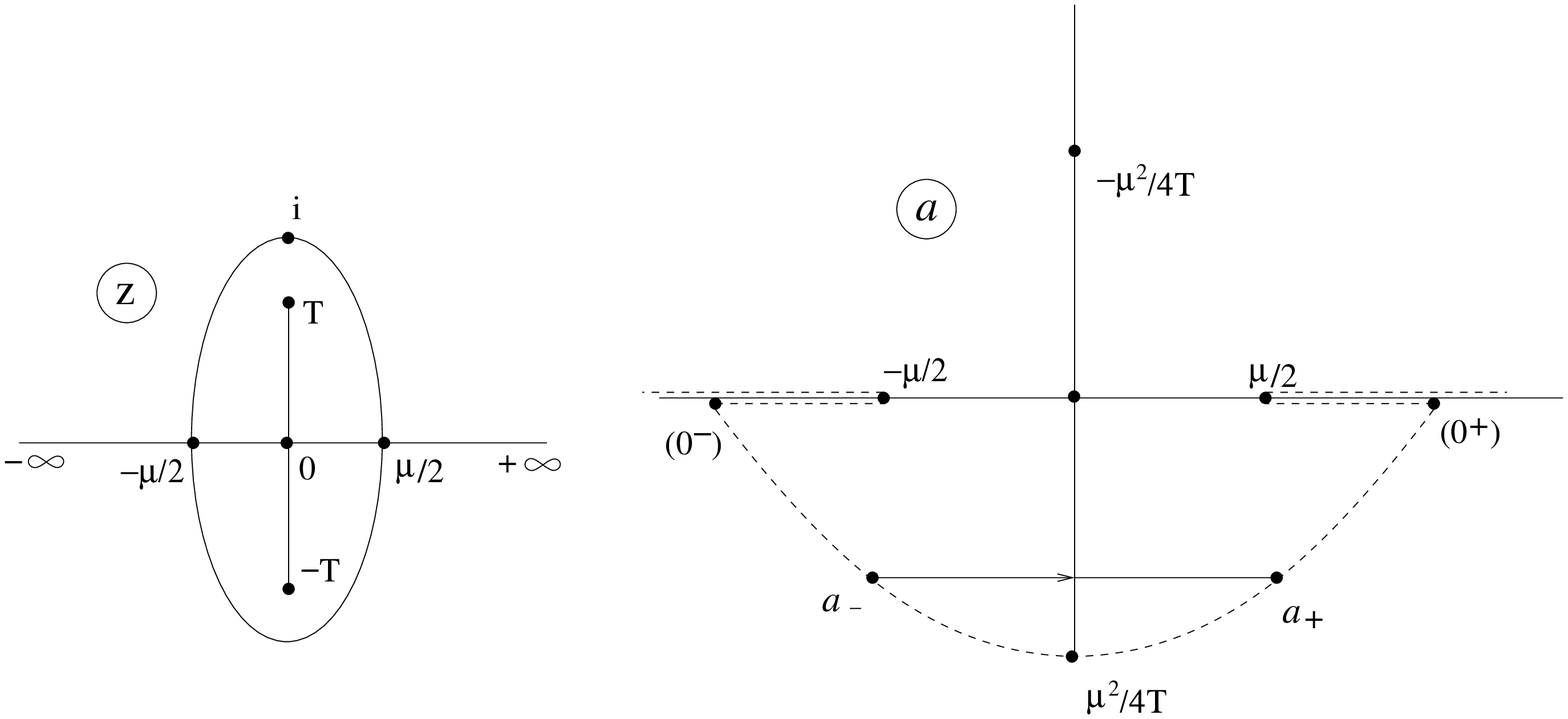}}
\caption{Mappings \eqref{mapaz} in the case $\m<2$. The dashed line
in the $a$-plane is the image of $\R$ by the map $\hat a(z)=a(x(z))$.
In particular,
$\hat a(\pm\infty)=\pm\infty,~~\hat a(\pm\mt)=\pm\mt,~~\hat a(0^\pm)=\pm\frac{\m}{2T},~~
\hat a( T)=\frac{\m^2}{4T}$. The solid line shows the contour of integration from
$a_-$ to $a_+$.} 
\label{mapam<}
\end{figure}

Let us now calculate $w(z)$,  $z\in\R$, for the case $\m>2$ (we only consider 
$|z|>T$) or for the case $\m<2$. Using \eqref{eqsmugen}, we obtain direct and inverse 
maps between the values of $a$ and $z$ by
\be\label{mapaz}
z=a+i\sqrt{1-\frac{4}{\m^2}a^2},~~~~a=\frac{\mt z -\sqrt{z^2-T^2}}{T^2}\cdot\mt  
\ee
The latter  maps ${\cal E^+}$ into a part of the lower $a$-halfplane and 
$\C^+\setminus{\cal E}$ into the upper $a$-halfplane, see Fig. \ref{mapam>},
Fig. \ref{mapam<}. Using \eqref{mapaz}, it is easy to check that if $z\ge T$, $\m>2$
or $z\ge 0$, $\m<2$,
then the corresponding $a=\hat a(z)=a(x(z))\in\R$ and $a>\mt$. Making change of 
variables $a(y)=\eta$ in 
\be \label{wzmgen1}
w(z)=\sign(\mt-z)\Im \int_{x(z)}^x\sqrt{(z-a(y))^2+b^2(y) }dy~
\ee 
and using \eqref{eqsmugen}, we obtain
\begin{align} \label{wzmgen2} 
w(z)=&\sign(\mt-z)\Im \int_{a(x(z))}^{a(x)}\sqrt{(z-\eta)^2+1-\frac{4\eta^2}{\m^2} }\cdot\frac{d\eta}{\mt\left(1-\frac{4\eta^2}{\m^2}\right)}\cr
=&\sign(\mt-z)\Im \int_{a(x(z))}^{a(x)}\sqrt{\msq(z-\eta)^2+\msq-\eta^2}
\cdot\frac{d\eta}{\msq-\eta^2}~.
\end{align}
Direct calculation show that the integrand is real-valued. Thus,
a contribution to the imaginary part of the latter integral can come
only from integration around the singular point $\eta=\mt$.
Repeating the previous argument (case $\m=2$), we obtain  
\be \label{wzmgenpm}
w(z)=\pt(\mt\mp z)~
\ee
if $\sign z=\pm 1$ respectively. Using the above technique, one could check
that in the case $\m>2$, $w(z)$ should be constant on the 
segment $[-T,T]\subset \R$. 
Thus, $w(z)$ satisfies the inequalities \eqref{ineq2} for all $\m>0$. 
In the pure radiational case $\m> 2$,  the proof of the sign distribution
\eqref{ineq1} is the same as for the case of $\m=2$ in Section \ref{exam1}.
This proof can be extended to the case  $m<2$ providing that the contour $\g$
does not intersect the branchcut of $f_0(z)$. However Lemma 4.11 from \cite{TVZ1}
shows that $\g_m=\g_m(x)$, which is defined by $f_0(z)$ given by \eqref{f1}, 
intersects the branchcut $[0,T]$ if $x$ is negative with sufficiently large $|x|$.
In this case one can use Corollary \ref{qeven} from Section \ref{symmetry} to show 
that $q_0(x,0,\e)$ is $O(\e)$ close to $\tilde q(x,0,\e)$ for all $x\in \R$.

wzmgenpm

Let us now calculate the jump 
\be\label{delfz0}
\D f(z)=f(x,z_-)-f(x,z_+)=h(x,z_+)-h(x,z_-)~
\ee
for the case $0<\mu< 2$, where the points $z_\pm$ are equal  but located
 on the opposite shores of the oriented vertical segment $[T,0]$.
According to \eqref{hxz}, we have
\be\label{delfz1}
\D f(z)=\int_{x(z_-)}^{x(z_+)}\sqrt{(z-a(y))^2+b^2(y)}dy
\ee  
where $x(z_-), x(z_+)$ are symmetrically located points on the dashed curve,
Fig. \ref{Bscr<}. Using the change of variables $\eta=\tanh x$, we obtain
\be\label{delfz2}
\D f(z)=\int_{\eta_-}^{\eta_+}\sqrt{z^2-\m z \eta+1+T^2\eta^2}\cdot\frac{d\eta}{1-\eta^2}~,
\ee
where, according to \eqref{mapaz},
\be\label{etapm}
\eta_\pm=\frac{\mt z \pm\sqrt{z^2-T^2}}{T^2}=\frac{\mt iy \pm\sqrt{-y^2-T^2}}{T^2}  
\ee
with $z=iy$, see Fig. \ref{mapam<}, where $a_\pm=\mt \eta_\pm$.

Observing that $\eta_\pm$ are zeroes of the radical in \eqref{delfz2}, we obtain 
\be\label{delfz3}
\D f(z)=\hf\int_\l\sqrt{z^2-\m z \eta+1+T^2\eta^2}\cdot\frac{d\eta}{1-\eta^2}~,
\ee
where $\l\subset\C^-$ is a closed, clockwise oriented  curve that contains the segment
$ [\eta_-,\eta_+]$. Thus,
\begin{align}\label{delfz4}
&\D f(z)=-\ipt\left[\Res\sqrt{z^2-\m z \eta+1+T^2\eta^2}\left|_\infty 
\pm \Res\sqrt{z^2-\m z \eta+1+T^2\eta^2}\right|_{\eta=\pm 1}\right] \cr
&=\ipt\left[\sqrt{z^2-\m z +\msq}- \sqrt{z^2+\m z +\msq}-2T\right]=i\pi(z-T) 
~.
\end{align}
Thus, we obtained the same value of $\D f(z)$ as \eqref{f1} has. It is now possible to reconstruct
$f_0(z)$, given by \eqref{f1} with $\m<2$,  from $w(z)$ defined by \eqref{wzmgenpm} and
$\D f(z)$  defined by \eqref{delfz4} on $[0,T]$.

\subsection{The ``Y'' - shape of Bronski}\label{examBr}

In \cite{Bronski1}, J. Bronski studied numerically  discrete spectrum of ZS problem
\eqref{ZS} with the potential $q(x,0\e)$ given by \eqref{ID}, where 
\be\label{bronpot}
A(x)=\sech 2x,~~~~~~~~~~S(x)=\m \sech 2x.~
\ee 
He found that for real valued $q(x,0\e)$, i.e., for $\m=0$,
 the accumulation curve for discrete eigenvalues of
\eqref{ZS}, \eqref{bronpot} in $\C^+$ in
the limit $\e\ra 0$ is a segment $[0,i]$ of the imaginary axis. In the case $\m=1$,
the accumulation curve has ``Y'' - shaped form, with $z=0$ located at the bottow of ``Y''.
The change of shape of the accumulation curve happens at the critical value 
$\m^*=2^{-\frac{3}{2}}$. In this section we show that the endpoints of the accumulation 
curve coicide with the branchpoints of $f_0(z)$, obtained from the potential \eqref{ID},
\eqref{bronpot} by the AH transformation \eqref{xtof}. In particular, $\m^*$ is a critical
value when one  branchpoint of $f_0(z)$ splits  into two.

For potential \eqref{ID}, \eqref{bronpot}, we have
\be\label{alpbron}
\a(x)=\frac{\m \sinh 2x}{\cosh^2 2x} + \frac{i}{\cosh 2x}=\frac{\m\sinh 2x+i\cosh 2x}{cosh^2 2x}~.
\ee
Then 
\be\label{alp'bron}
\a'(x)=2\m \frac{\cosh^2 2x- 2 \sinh^2 2x}{\cosh^3 2x} -2i \frac{\sinh 2x}{\cosh^2 2x}
=-\frac{\m \cosh 4x+i\sinh 4x-3\m}{\cosh^3 2x}~,
\ee
so, after some algebra, equation $\a'(x)=0$ for the branchpoints can be written as
\be\label{brpteq} 
(\m+i)e^{8x}-6\m e^{4x}+\m-i=0~.
\ee 
Solution of  \eqref{brpteq} is given by 
\be\label{solrpteq}
\left(e^{4x} \right)_{1,2}=\frac{3\m\pm\sqrt{8\m^2-1}}{\m+i}~. 
\ee
Thus, for $\m=0$ we obtain $e^{4x}=1$, so $x=0$ is a branchpoint and $z=\a(0)=i$
is the corresponding log point in the spectral plane. 

The critical value  $\m^*$
is given by equation $8\m^2-1=0$, which yields $\m^*=2^{-\frac{3}{2}}$. 
For $\m\le \m^*$, direct calculation show that $|e^{4x}|=1$ for solution $e^{4x}$ given by  \eqref{solrpteq} with the positive sign. Thus, the corresponding log point $x\in\i R$.
Substituting $\m^*$ into \eqref{solrpteq}, we find 
$e^{4x^*}=\frac{1}{3}(1-2\sqrt{2}i)$, where 
$x^*=-\frac{i}{2} \tan^{-1}(2\sqrt{2})$ is the the branchpoint that corresponds to $\m^*$. 
Direct calculations show that the corresponding log point on the spectral plane (the double point on Fig. 11 in \cite{Bronski1}) is given by $z^*=\a(x^*)=i\frac{3\sqrt{3}}{4\sqrt{2}}\thickapprox 0.91856 i$. 

To calculate the log point(s) (endpoint(s) of the accumulation curve) for other values of $\m$,
we rewrite \eqref{brpteq} as 
\be\label{brpteq1} 
(\m^2+1)\cosh^2 4x - 6\m^2\cosh 4x + 9\m^2 -1=0~,
\ee 
which yields
\be\label{solrpteq1}
\left(\cosh 4x \right)_{1,2}=\frac{3\m^2 \pm\sqrt{1-8\m^2}}{\m^2+1}~. 
\ee
Substituting $\m=0$, we see that for $\m<\m^*$ we need to choose the positive sign in \eqref{solrpteq1}.
Calculating  now $\cosh 2x$ and $\sinh 2x$, we can express logpoints $\a(x)$ for an arbitrary $\m>0$
as 
\be\label{logbron}
\a(x)=\frac{-\m\sqrt{2\m^2-1\pm\sqrt{1-8\m^2}}+i\sqrt{4\m^2+1\pm\sqrt{1-8\m^2}}} 
{4\m^2+1\pm\sqrt{1-8\m^2}}\cdot\sqrt{2(\m^2+1)}~,
\ee
were only the plus sign should be used for $\m<\m^*$ (note the choice of signs for the radicals 
representing $\cosh 2x$ and $\sinh 2x$).

Expression \eqref{logbron} 
provides  the  endpoints of the two ``legs'' of the  ``Y''-shaped accumulation curve
(as $\e\ra 0$) for the points of the discrete spectrum  in the case $\m>\m^*$. In the case
$0\le \mu\le\m^*$, it gives the tip of the vertical segment on the imaginary axis where the points 
of the discrete spectrum accumulate.

\subsection{Double hump initial data}\label{exam2}

Another example of an interesting  initial data is
\be\label{dhumpdata}
a(x)=\tanh x,~b(x)=\sech x-k\sech^2 x,~~~~k\in [0,1],
\ee
which contains double hump cases. Indeed, if 
 $k>\hf$, then $b=u-ku^2$,
where $u=\sech x$, has a global maximum at $u=1/2k\in(0,1)$. So,
$x=\pm\cosh^{-1}(2k)$ are two points of maximum of $b(x)$.

The points of ramification of the map 
\be\label{alpdub}
\a(x)=\tanh x+i( \sech x -k\sech^2 x)~~~~{\rm with}~~~~\a'(x)=\frac{1-i\sinh x +2ik\tanh x}{\cosh^2 x}
\ee
satisfy equation
\be\label{alpdub'=0} 
1-i\sinh x +2ik\tanh x=0.
\ee
Coefficients of this equation are $2\pi i$ periodic functions. The substitution
$u=\sinh x$ in \eqref{alpdub'=0} yields a forth order polinomial equation in $u$.
Thus, there are no more than 4 ramification points of $\a(x)$ within a horizontal 
strip of width $2\pi$ in the complex $x$-plane.

\begin{figure}
\centerline{
\includegraphics[height=5cm]{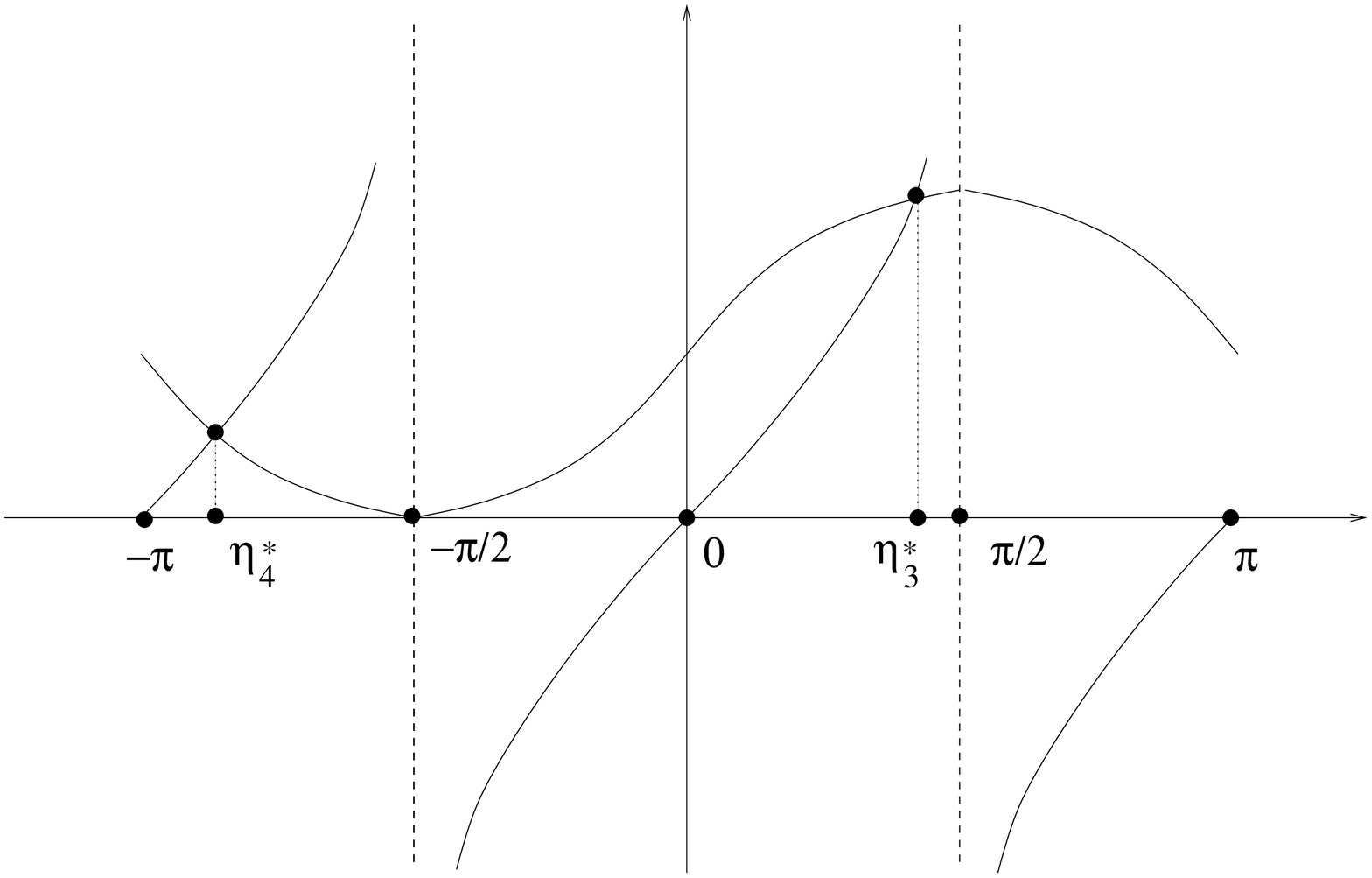}}
\caption{Intersections of functions $1+\sin\eta$  and $2k\tan\eta$ and points $\eta^*_{3,4}$.} 
\label{doubfg1}
\end{figure}

Separating real and imaginary parts of \eqref{alpdub'=0}, we obtain the system
\be\label{rptsys}
\begin{cases}
&\hf\sinh(2\x)\cos(2\eta)-2k\sinh \x\cos\eta-\sinh\x\sin\eta=0 \cr
&\hf\cosh(2\x)\sin(2\eta)-2k\cosh\x\sin\eta+\cosh \x\cos\eta=0, \cr
\end{cases}
\ee
where $x=\x+i\eta$. The first equation in \eqref{rptsys} has a common factor  $\sinh \x$.
Setting it zero, we obtain $\x=0$. Then the second equation \eqref{rptsys} becomes
 $1+\sin \eta=2k\tan \eta$. Fig. \ref{doubfg1}
shows that this equation has one positive  solution $\eta_3^*$ 
and one negative solution $\eta^*_4$ on $(-\pi,\pi)$, where $0<\eta^*_3<\pt$
and $\eta^*_4<-\pt$. We denote by $x_{3,4}^*$ the corresponding ramification points in
the complex $x$-plane.

Considering $\x\neq 0$, we rewrite the first equation in \eqref{rptsys} as
\begin{equation}\label{rptsys1}
\cosh\x\cos(2\eta)-2k\cos\eta-\sin\eta=0 ,
\end{equation}
which yields
\be\label{chxi0}
\cosh\x=\frac{\sin\eta+2k\cos\eta}{\cos(2\eta)}~.
\ee
Substitution of \eqref{chxi0} into the second equation of \eqref{rptsys} yields
\be\label{ch^2xi}
\cosh^2\x=\hf\frac{(4k^2-1)\sin(2\eta)-2k\cos(2\eta)+\hf\sin(2\eta)\cos(2\eta)}{\sin(2\eta)\cos(2\eta)}~.
\ee
From \eqref{chxi}-\eqref{ch^2xi} after some algebra we obtain 
\be\label{cubic}
u^3+4k^2u+4k=0,
\ee
where $u=\sin(2\eta)$.
Equation \eqref{cubic} should yield the remaining two points of ramification. In the case $k=\hf$
equation \eqref{cubic} becomes $(u+1)[u^2-u+2]=0$, which has the only real root $u=-1$.
Substituting the corresponding $\eta=-\frac{\pi}{4}$ into the second equation of \eqref{rptsys},
we obtain $\cosh 2\x=2\sqrt{2\cosh\x}$, or $\x=\pm\cosh^{-1}\left(1+\frac{1}{\sqrt{2}}\right)$.
Thus, for $k=\hf$ we obtain the remaining ramification points
\be \label{xk=hf}
x^*_{1,2}=\pm\cosh^{-1}\left(1+\frac{1}{\sqrt{2}}\right)-\frac{i\pi}{4}~.
\ee

It is clear that in the case of arbitrary $k\ge 0$ equation \eqref{cubic} has only one real solution.
Indeed, if that is not the case, then \eqref{cubic} would have a multiple real root for some
$k>0$. But equation $m'(u)=3u^2+k^2=0$, where $m(u)= u^3+4k^2u+4k$ has no real roots, so there is a unique real root $u(k)$ of \eqref{cubic}. Moreover, $u(k)\in (0,-1]$, 
since $m(0)>0$ and $m(-1)\le 0$. Thus, $\eta=\hf\sin^{-1}u(k)$ or $\eta=-\pt-\hf\sin^{-1}u(k)$.
Substituting one of these values into \eqref{chxi0} (the former if $k\ge\hf$ and the latter
if $k\le\hf$), we find the real components of the two remaining points of ramification $x_{1,2}^*$.
Note that $\Re x_1^*=-\Re x_2^*$.

\begin{figure}
\centerline{
\includegraphics[height=5cm]{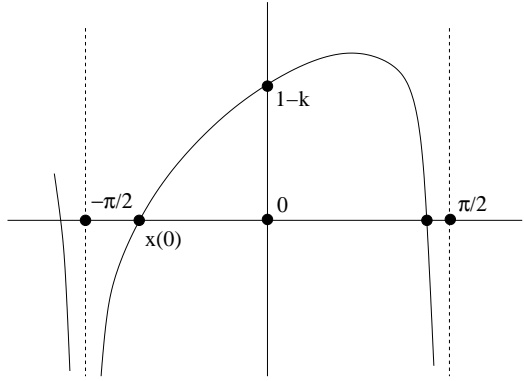}}
\caption{The graph of $\frac{1+\sin\eta}{\cos\eta}-\frac{k}{\cos^2\eta}$.} 
\label{doubfg2}
\end{figure}

To calculate the pre-images of $z=0$ of the map \eqref{alpdub}, we set equation
\be\label{zeq}
z\cosh^2x=\sinh x \cosh x+i\cosh x -ik~.
\ee
For the same reasons as above, for every $z\in\overline{\C^+}$ this equation has four roots
in the strip $-\pi<\Im x\le \pi$. Assuming $z\in\R$ and
separating real and imaginary parts of \eqref{zeq}, we obtain 
\be\label{zsys}
\begin{cases}
&z(\cosh(2\x)\cos(2\eta)+1)=\sinh(2\x)\cos(2\eta)-2\sinh\x\sin\eta\cr
&z(\sinh(2\x)\sin(2\eta)=\cosh(2\x)\sin(2\eta)+2\cosh\x\cos\eta-2k~.
\end{cases}
\ee
If $z=0$ then the first equation yields: 1) $\sinh\x=0$, or; 2)
\be\label{z=02}
\cosh\x\cos(2\eta)=\sin\eta
\ee 
Substituting $\x=0$ in the second equation \eqref{zsys} yields
\be\label{z=0im}
\cos\eta(1+\sin\eta)=k,
\ee
which has exactly one positive and one negative root on $(-\pt,\pt)$.
The latter root is denoted $x(0)$, see Fig. \ref{doubfg2}

To analize the remaining roots of \eqref{zeq} with $z=0$, we  substitute 
\be\label{chxi}
\cosh\x=\frac{\sin\eta}{\cos 2\eta}
\ee
into the second equation \eqref{zsys} with $z=0$. After some algebra, we obtain
\be\label{z0comp}
\sin^32\eta+2k\sin^22\eta-2k=0~.
\ee
Introducing $u=\sin2\eta$, we can rewrite \eqref{z0comp} as
\be\label{m(u)}
m(u)=u^3+2ku^2-2k=0~.
\ee
It is easy to see  that $m(u)$ has a real root $u(k)\in[0,1)$. Let us show that
$m(u)$ has only one real root. Indeed, $m'(u)=(3u+4k)u$. So, critical points
of $m(u)$ are $u_1=-\frac{4k}{3}$ and $u_2=0$. It is clear that $u_1$ is a local 
maximum, $u_2$ is a local minimum and $m(u_2)<0$. Since 
$m(u_1)=\left( \frac{32}{27}k^2-2\right)k\le 0$ for all $k\in[0,1]$,
we conclude that  $m(u)$ has a unique real root $u(k)$ and $u(k)\in[0,1)$.

\begin{figure}
\centerline{
\includegraphics[height=5cm]{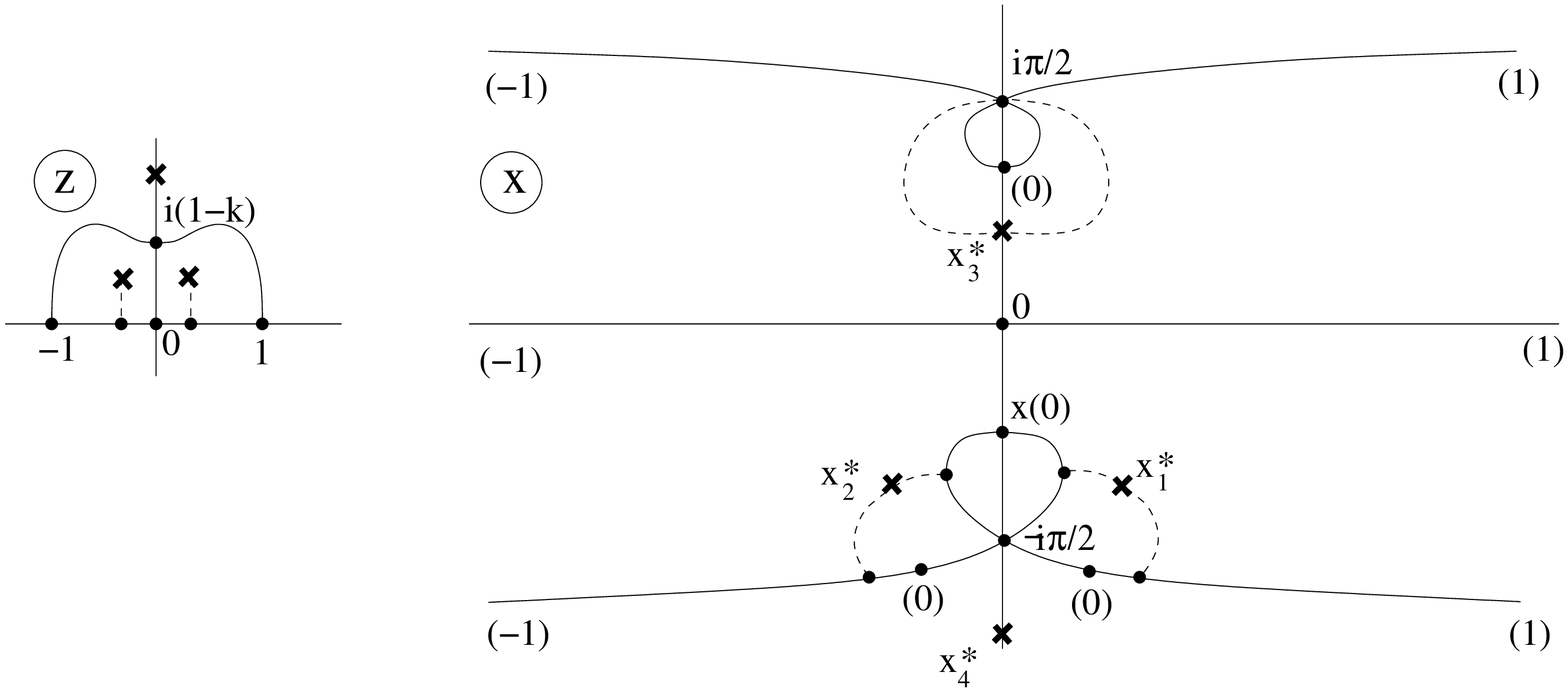}}
\caption{The image $\Cscr$ and the domain $\Bscr$ of the map $z=\a(x)$, where $\a(x)$ is defined by \eqref{alpdub}.
Ramification  points $x^*_i$, $i=1,2,3,4$ in the $x$-plane  and the  corresponding log points in $\Cscr$  are marked
by ``crosses".  Vertical cuts in $\Cscr$  and their   pre-images  in $\Bscr$ are shown by dashed lines.  Round brackets
in the $x$-plane are used to denote pre-images of points. The region $\Cscr$ consists of the upper $z$-halfplane with
three cuts.  The upper boundary of $\Bscr$ consists of two curves, connecting the pre-images of $\pm 1$ with $\ipt$
(pre-images of $z>1$ and $z<-1$ respectively), and the closed dashed curve (the pre-image of the cut from $\a(x_3^*)$
to $i\infty$).  The lower boundary of $\Bscr$ consists of two dashed curves that contain points $x_1^*, x_2^*$
(pre-images of verical cuts in ${\cal E}^+$),   the curve that connects them and passes through $x(0)$, and curves
connecting the dashed curves with $\pm \infty$ (the latter three curves form a pre-image ob the interval $(-1,1)$).}

\label{calBdoub}
\end{figure}

Thus, we obtain $\sin 2\eta\in [0,1)$, which means $2\eta\in [0,\pi]$ or 
$2\eta[-2\pi,-\pi]$. That means $\eta\in [0,\pt]$ or $\eta\in[-\pi,-\pt]$.
However, according to \eqref{chxi}, we have a restriction
\be\label{ch>1}
\frac{\sin\eta}{\cos 2\eta}\ge 1, 
\ee
which imply that 
\be\label{sinetaineq}
\sin \eta<-\frac{1}{\sqrt{2}}~~~~~{\rm or}~~~~~ \hf\le \sin \eta <\frac{1}{\sqrt{2}}~.
\ee
Combining \eqref{sinetaineq} with the above restrictions on $\eta$, we obtain
\be\label{etaineq}
 -\frac{3\pi}{4}<\eta <-\frac{\pi}{4}~~~~~{\rm or}~~~~~\frac{\pi}{6}\le\eta<\frac{\pi}{4}.
\ee
However, we will show that latter option is not possible. Indeed, implicitly
differentiating  \eqref{m(u)}, we obtain 
$$\frac{du}{dk}=\frac{2(1-u^2)}{(3u+4k)u}>0$$
for any $u\in(0,1)$. Thus, the root $u(k)$ is monotonically increasing.
Substituting $u(k)=\sin 2\eta=\sin\frac{\pi}{3}$ into \eqref{m(u)},
we obtain the corresponding $k=\frac{3\sqrt{3}}{4}>1$. Thus, for all $k\in[0,1]$,
the corresponding $u(k)<\frac{\sqrt{3}}{2}$, so that \eqref{ch>1} cannot be satisfies 
for the corresponding $\eta$.

Thus, we proved that for any $k\in[0,1]$ 
there exists a unique $\eta$ and unique $\cosh\xi$,
connected through \eqref{chxi}, that satisfy \eqref{zsys}. Moreover,
such $\eta\in\left( -\frac{3\pi}{4},-\frac{\pi}{4}\right) $.  Thus, we proved
existence of the second pair of roots of $\a(x)=0$ located in the strip
$ -\frac{3\pi}{4}<\Im x<-\frac{\pi}{4}$ and symmetrical with respect
to the imaginary axis. 

Obtained information allows us to sketch the domain $\Bscr$
for the map \eqref{dhumpdata}, see Fig. \ref{calBdoub}.
Calculation of $w(z)$ and of the jumps $\D f_0(z)$ over the branchcuts is not
included into this paper.

\end{document}